\documentclass{jfm}

\usepackage{graphicx}
\usepackage{subcaption}
\usepackage{xcolor}
\usepackage{natbib}
\usepackage{amsmath}
\usepackage{amssymb}
\usepackage{bm}
\usepackage{hyperref}
\usepackage{cleveref}

\title[Numerical study of filament suspensions at finite inertia]
{Numerical study of filament suspensions at finite inertia}

\author[A. Alizad Banaei, M. E. Rosti and L. Brandt ]{Arash Alizad Banaei, Marco Edoardo Rosti and Luca Brandt}

\affiliation{Linn\'{e} Flow Centre and SeRC (Swedish e-Science Research Centre), \\KTH Mechanics, SE 100 44 Stockholm, Sweden}

\begin{document}

\maketitle

\begin{abstract}
We present a numerical study on the rheology of semi-dilute and concentrated filament suspensions of different bending stiffness and Reynolds number, with the immersed boundary method used to couple the fluid and solid. The filaments are considered as one-dimensional inextensible slender bodies with fixed aspect ratio, obeying the Euler-Bernoulli beam equation. 
To understand the global suspension behavior we relate it to the filament microstructure, deformation and elastic energy and examine the stress budget to quantify the effect of the elastic contribution.
At fixed volume fraction, the viscosity of the suspension reduces when decreasing the bending rigidity and grows when increasing the Reynolds number. The change in the relative viscosity is stronger at finite inertia, although still in the laminar flow regime as considered here. Moreover, we find the first normal stress difference to be positive as in polymeric fluids, and to increase with the Reynolds number;  its value has a peak for an intermediate value of the filament bending stiffness. The peak value is found to be proportional to the Reynolds number, moving towards more rigid suspensions at larger inertia. 
Moreover, the viscosity increases when increasing the filament volume fraction, and the rate of increase of the filament stress with the bending rigidity is stronger at higher Reynolds numbers and reduces with the volume fraction. We show that this behaviour is associated with the formation of a more ordered structure in the flow, where filaments tend to be more aligned and move as a compact aggregate, thus reducing the filament-filament interactions despite their volume fraction increases.
\end{abstract}

\section{Introduction}
\subsection{Motivations and objectives}
Filaments suspensions are found in many applications, such as material reinforcement, pulp and paper industry, they are relevant to the swimming of microorganisms and can induce drag reduction in turbulent flows. In particular, the study of the rheology of fiber suspensions is essential in many industrial applications, such as paper production and composite materials \citep{lundell2011fluid,lindstrom2008simulation}. Suspensions of fibers are characterized by a complex rheology which is affected by a large number of parameters, such as the fiber aspect ratio and mass density, their deformability and volume fraction, and not least the flow inertia. The majority of previous studies focused mainly on the effect of two of these parameters, i.e.,~the fiber aspect ratio and volume fraction, in the limit of vanishing inertia. In addition, only few experimental studies considered deformable fibers \citep{keshtkar2009rheological,kitano1981rheology}. Numerical simulations have been used to address fiber suspensions only quite recently; in these studies the effects of deformability are often accounted for by modeling the fibers as chains of connected spheres or cylinders \citep{joung2001direct,wu2010numerical}. In this work, we study study the rheology of semi-dilute and concentrated suspensions of flexible fibers, modeled as continuously deformable objects. A wide range of flexibilities and Reynolds numbers will be considered, thus including inertial effects withinin the limit of the laminar flow regime.

\subsection{Suspensions of rigid fibers}
The rheology of rigid fiber suspensions has been extensively studied in the past both experimentally and numerically. Typically, fiber suspensions are characterized by their number density, $\frac{nL^3}{V}$ where $\frac{n}{V}$ is the number of fibers per unit volume and $L$ their length. Three regimes are thus identified: dilute, semi-dilute and concentrated suspensions. In the dilute limit, $\frac{nL^3}{V}<1$, fiber-fiber interactions are negligible and fibers move independently from each other. In the semi-dilute regime, $1<\frac{nL^3}{V}<\frac{L}{d}$ with $d$ the fiber diameter, fiber-fiber interactions start to affect the global dynamics, and finally in the concentrated regime, $\frac{nL^3}{V}>\frac{L}{d}$, interactions between fibers are dominant \citep{wu2010numerical}.

\cite{blakeney1966viscosity} measured the effect of the solid volume fraction of nylon fibres on the suspension viscosity in the dilute regime, with concentrations up to $1\%$. It was found that the relative viscosity rapidly grows for volume fractions above the critical value of $0.42\%$, then slightly decreases for volume fractions between $0.5\%$ and $0.6\%$ followed by a second rapid increase for volume fractions above $0.6\%$. \cite{bibbo1987rheology} experimentally investigated the rheology of semi-concentrated rigid fibers suspensions in both Newtonian and Non-Newtonian solvents, and observed that the relative viscosity is only a function of the volume fraction and independent of the fiber aspect ratio for large enough values of the imposed shear rate for a Newtonian suspending fluid. Similar experiments were performed more recently by \cite{chaouche2001rheology} and \cite{djalili2006fibre}. The former authors found a nearly Newtonian behaviour in semidilute suspensions, while shear-thinning was observed in more concentrated regimes; also, this non-Newtonian behaviour was found to increase with the fiber concentration and to decrease with the solvent viscosity. \cite{djalili2006fibre} observed a strong dependency of the suspension viscosity on the fiber aspect ratio for volume fractions above $5\%$ due to the presence of friction forces during fiber-fiber interactions.

Numerical simulations of fiber suspensions have been performed only quite recently. \cite{yamane1994numerical} were the first to study dilute suspensions of non-Brownian fibers under shear flow by exploiting analytical solutions for rigid slender bodies: they considered short range interactions between fibers due to lubrication forces but neglected long range interactions. These authors concluded that the relative viscosity of the suspension is only slightly altered by fiber-fiber interactions in this dilute regime. \cite{mackaplow1996numerical} considered fibers as line distributions of Stokeslets and used slender body theory to determine the fiber-fiber interactions; they observed  that the suspension viscosity can be well predicted analytically considering simple two-body interactions for dilute and semidilute concentrations. \cite{lindstrom2008simulation} performed numerical simulations of rigid fiber suspensions to study fiber agglomeration in the presence of friction forces. These authors observed that the apparent viscosity increases non-linearly with the friction coefficient and fibers tend to flocculate even in the semidilute regime. The role of the fiber curvature on the effective viscosity of suspensions of rigid fibers was studied by \cite{joung2002viscosity} who showed that this results in a large increase of the suspension viscosity already for small curvatures.

\subsection{Suspensions of flexible fibers}
While most of the previous studies on rigid fiber suspensions consistently report an increase of the suspension viscosity with the volume fraction,  differing results have been reported in the past on the effect of the fiber flexibility on the global suspension rheology. One of the first study on flexible fibers was the experimental investigation by \cite{kitano1981rheology}. These authors considered vinylon fibers immersed in a polymeric liquid and observed an increase of the suspension viscosity and of the first normal stress difference with the volume fraction and fiber aspect ratio. Although these authors mentioned that the fiber deformability may affect the rheological properties of the suspension, its effect was not discussed explicitly. This was done more recently by \cite{keshtkar2009rheological} who investigated fibers suspensions with different flexibilities and high aspect ratios in Silicon oil. These authors found that the viscosity of the suspensions increases when the fiber is deformable. \cite{yamamoto1993method} proposed a numerical method to simulate flexible fibers by modeling them as chains of rigid spheres joined by springs, which allow each element to stretch, bend and twist. \cite{joung2001direct} used this method and found an increase of the suspension viscosity with the fiber elasticity. A similar procedure was adopted by \cite{schmid2000simulations} who modeled the flexible fibers as chain of rods connected by hinges. Using this method, \cite{switzer2003rheology} studied flexible fiber suspensions and found that the viscosity of the suspension is strongly influenced by the fiber equilibrium shape, by the inter-fiber friction, and by the fiber stiffness. In particular, they reported a decrease of the relative viscosity with the ratio of the shear rate to the elastic modulus of the fibers. Finally, the rod-chain model was also used by \cite{wu2010method,wu2010numerical} who found again an increase of the suspension viscosity with the fibers flexibility, in contrast with the computational results by \cite{switzer2003rheology} who employed the same rod-chain model for fibers with aspect ratio of $75$ and the experimental results by \cite{sepehr2004rheological} who studied suspensions of fibers with aspect ratio $20$ in viscoelastic fluids. Note that, suspensions of other deformable object, such as particles of viscoleastic material and capsules (thin elastic membranes enclosing a second liquid) also exhibit a suspension viscosity decreasing with elasticity and deformation \citep{matsunaga2016rheology,rosti2018suspensions}. In particular, \cite{rosti2018rheology} and \cite{rosti2018suspensions} show that the effective suspension viscosity can be well predicted by empirical fits obtained for rigid particle suspensions if the deformability is taken into account as a reduced effective volume fraction.

\subsection{Outline}
In this paper we focus on the effect of finite inertia and flexibilty on the rheological properties of flexible fiber suspensions. We perform numerical simulations and assume the fibers as continuous flexible slender bodies obeying the Euler-Bernoulli beam theory; the fiber dynamics is coupled to the fluid equation by an Immersed Boundary Method. Note that, the fibers are inextensible, their lengths remaining constant during the deformation. The paper is outlined as follows. Section 2 describes in details the governing equations of both the fluid and solid phases and the numerical method used to solve this multiphase problem; we also provide three validation cases by comparing our results with those in the literature. In section 3 we present the flow configurations under investigation while the results are presented in section 4. In particular, we perform a parametric study varying the flow Reynolds number, the fiber bending rigidity and the volume fraction and examine the suspension viscosity, the first normal stress difference and the elastic energy of the fibers suspension. Finally, we end the manuscript by summarizing the main conclusions of the work.

\section{Governing equations and numerical method}
\subsection{Flow field equations}
We consider an incompressible suspending fluid, governed by the Navier-Stokes equations. In an inertial, Cartesian frame of reference the non-dimensional momentum and mass conservation equations for an incompressible flow read
\begin{equation} \label{eq:NSm}
  \frac{\partial \bm{u}}{\partial t}
  + \bm{\nabla} \cdot \left( \bm{u} \otimes \bm{u} \right) = 
  - \bm{\nabla} p 
  + \frac{1}{Re} \nabla^2 \bm{u} + \bm{f},
\end{equation}
\begin{equation} \label{eq:NSc}
  \bm{\nabla} \cdot \bm{u} = 0,
\end{equation}
where $\bm{u}$ is the velocity field, $p$ the pressure, $\bm{f}$ a volume force (used to account for the suspended filaments), and $Re=\rho \dot{\gamma} {L^*}^2/\mu$ the Reynolds number where $\rho$ is the fluid density, $\mu$ the fluid dynamic viscosity, $L^*$ the reference length scale, here the filament length, and $\dot{\gamma}$ the applied shear rate. 

\subsection{Filament dynamics}
In the present framework, we study neutrally buoyant and inextensible filaments. The dynamics of a thin flexible filament can be described by the Euler-Bernoulli equation \cite[see e.g.][]{segel_2007a} under the constraint of inextensibility, which reads as
\begin{equation} \label{Eul_Ber}
\left(\frac{ \rho_f/A_f-\rho}{\rho}\right) \frac{\partial^2 \bm{X}}{\partial t^2} = \frac{\partial}{\partial s} \left( T \frac{\partial \bm{X}}{\partial s}\right) - B \frac{\partial^4 \bm{X}}{\partial s^4}+\frac{\Delta \rho}{\rho_f}Fr\frac{\textbf{g}}{g} - \bm{F} + \bm{F}^f , 
\end{equation}
\begin{equation} \label{inext}
\frac{\partial \bm{X}}{\partial s} \cdot \frac{\partial \bm{X}}{\partial s} = 1,
\end{equation}
where $\rho_f$ is the filament linear density (mass per unit length) and $A_f$ their cross-sectional area. $\bm{X}$ is the filament position, $s$ the curvilinear coordinate along the filaments, $T$ the tension, $B=\frac{EI}{\rho_f\dot{\gamma}^2 {L^\ast}^4}$ the bending rigidity with $E$ the elastic modulus and $I$ the second moment of area for filament cross section, $\bm{F}$ the fluid-solid interaction force per unit length, $\bm{F}^f$ the force used to model the interactions between adjacent filaments and walls. Finally, $\Delta \rho$ is the linear density difference between the filaments and the surrounding fluid and $Fr=\frac{g}{L^\ast\dot{\gamma}^2}$ is the Froude number with $\textbf{g}$ the gravitational acceleration vector and $g=|\textbf{g}|$. Since we are studying neutrally buoyant filaments ($\Delta \rho=0$), the gravity term is null as well as the left hand side (LHS) of (\ref{Eul_Ber}), which therefore requires a specific numerical treatment as detailed below. Equation (\ref{Eul_Ber}) is made non-dimensional with the following characteristic scales: $L^*$ for length, $\dot{\gamma}^{-1}$ for time, $\rho_{f}L^{*2}\dot{\gamma}^2$ for tension and $\rho_{fl}L^{*4}\dot{\gamma}^2$ for the fluid-solid interaction and repulsive forces. As the filaments are suspended in the fluid, we impose at the two free ends zero torque, force and tension, i.e. 
\begin{equation}
\frac{\partial^2{\bm{X}}}{\partial{s}^2}=0,\;\;\;\;\;\;\;\;\;\frac{\partial^3{\bm{X}}}{\partial{s}^3}=0,\;\;\;\;\;\;\;\;\;T=0.
\end{equation}

\subsection{Numerical method}
\subsubsection{Immersed boundary method}
The fluid-solid coupling is achieved by an Immersed Boundary Method (IBM), as first proposed by \cite{peskin1972flow} to study the blood flow inside the heart. The main feature of the IBM is that the numerical grid does not need to conform to the geometry of the object, which is instead represented by a volume force distribution $\bm{f}$ that mimics the effect of the object on the fluid, typically the no-slip and no-penetration boundary conditions at the solid surface. In this approach, two sets of grid points are needed: a fixed Eulerian grid $\bm{x}$ for the fluid and a moving Lagrangian grid $\bm{X}$ for the flowing deformable structure. The volume force arising from the action of the filaments on the fluid is obtained by the convolution onto the Eulerian mesh of the singular forces estimated on the Lagrangian nodes; these are computed using the fluid velocity interpolated at the location of the Lagrangian points \citep{rosti2018flexible,rosti_olivieri_banaei_brandt_mazzino_2019a}. These so-called interpolation and spreading operations are usually performed by means of regularized delta functions, in our case the one proposed by \cite{roma1999adaptive}.

 \begin{figure}
     \centering 
    \includegraphics[width = 0.95\textwidth,trim={0 9cm 0 0},clip]{{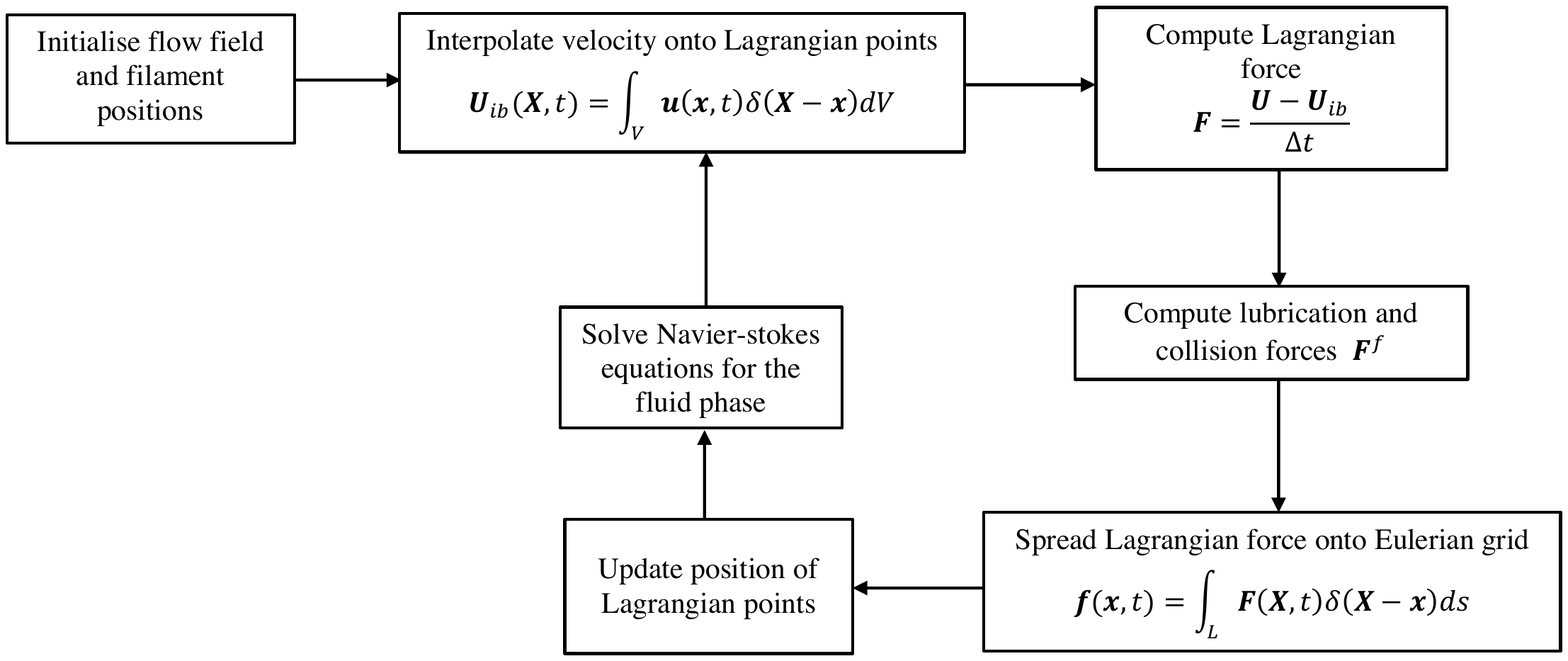}}
     \caption{Flowchart of the computational procedure used for the suspended filaments}
     \label{flowchart}
 \end{figure}

The computational procedure is depicted in figure \ref{flowchart}. At every time step, first, the fluid velocity is interpolated onto the Lagrangian grid points
\begin{equation} \label{interpolation}
\bm{U}_{ib}=\int_V \bm{u} \delta \left( \bm{X}-\bm{x} \right) dV,\
\end{equation}
where $\delta$ is the Dirac delta function \citep{roma1999adaptive}. These integral represents the interpolation from the Eulerian velocity field in a sphere with radius equal to $1.5\Delta x$ (where $\Delta x$ is the Eulerial grid spacing) to the Lagrangian point velocity. The fluid and solid equations are coupled by the fluid-solid interaction force
\begin{equation}\label{force}
\bm{F}=\frac{\bm{U}-\bm{U}_{ib}}{\Delta t},
\end{equation}
where $\bm{U}_{ib}$ is the interpolated velocity on the Lagrangian points defining the filaments, $\bm{U}$ the velocity of the Lagrangian points, and $\Delta t$ the time step. The Lagrangian force is then spread back to the fluid
\begin{equation}\label{spread}
\bm{f}=\frac{\pi}{4} r_P^2 \int_{L_f} \bm{F} \delta \left( \bm{X}-\bm{x} \right) ds,
\end{equation}
where $r_P=d/L^*$ is the filament aspect ratio and the factor $\frac{\pi}{4} r_P^2$ arises from dimensional arguments in the limit of mono-dimensional filaments. Equation (\ref{spread}) is used to represent the singular Lagrangian force of each point of the filament onto the Eulerian grid. \\
\subsubsection{Short-range interactions between the filaments}
We now discuss the forces which contribute to the interaction force between filaments, $\bm{F}^f$ which is decomposed as $\bm{F}^f = \bm{F}^l+\bm{F}^{lc}$ where $\bm{F}^{lc}$ is the lubrication correction and $\bm{F}^c$ is the contact force. In order to capture short range interactions between filaments whose distance is of the order of the numerical mesh size, we use the lubrication correction proposed by \cite{lindstrom2008simulation}. The model is based on the force between two infinite cylinders obtained for two different cases: when the two cylinders are parallel and when they are not. In the non-parallel case, \cite{yamane1994numerical} derived a first order approximation of the lubrication force,
\begin{equation} \label{yamane}
\bm{F}_1^{l}=\frac{-12 }{Re \sin{\alpha}} \frac{\dot{\bm{h}}}{h},
\end{equation}
where $h$ denotes the shortest distance between the cylinders and $\alpha$ the contact angle. To use this in the Euler-Bernoulli equations governing the filament dynamics, the force is converted into a force per unit length, i.e.,~it is divided by the Lagrangian grid spacing $\Delta s$. Equation (\ref{yamane}) cannot be used for the lubrication between parallel cylinders since ${\bm{F}_1^{l}\to\infty}$ as ${\alpha\to 0}$; in this case, a first order approximation of the force per unit length was derived by \cite{kromkamp2005shear}:
\begin{equation} \label{kromkamp}
\begin{aligned}
& \bm{F}_2^{l}=\frac{-4}{\pi Re r_p^2} \left(A_0+A_1\frac{h}{r}\right)\left(\frac{h}{r}\right)^{-3/2} \dot{\bm{h}},\\
        & A_0=3\pi \sqrt{2}/8, \ A_1=207\pi \sqrt{2}/160,       
        \end{aligned}
\end{equation}
where $r$ is the radius of the cylinders ($r=d/2$). Based on equations (\ref{yamane}) and (\ref{kromkamp}), the following approximation of the lubrication force for two finite cylinders can be assumed \citep{lindstrom2008simulation}:
\begin{equation} \label{lind}
\bm{F}^l=min(\bm{F}^l_{1}/\Delta s,\bm{F}^l_{2}).
\end{equation}
Finally, the lubrication force between walls and filaments is found by considering the walls as cylinders of infinite radius and assuming the contact area to be that between two cylinders with equivalent radius, i.e.,~$r_{eq}=r/2$ \citep{lindstrom2008simulation}. In our simulations, when the shortest distance between two Lagrangian point becomes lower than $d/4$, we impose the lubrication correction $\bm{F}^{lc}=\bm{F}^l-\bm{F}^l_0$, being $\bm{F}^l_0$ the lubrication force at a distance of $d/4$. We also performed some tests with an activation distance equal to $d$ and found that the change in the global suspension viscosity was less than 1.5\%. The total lubrication force acting on the $i$-th element of a filament is obtained as:
\begin{equation}\label{lubcorri}
\bm{F}^{lc}_{i}=\sum_{j \neq i}^{nl} \bm{F}^{lc}_{ij},
\end{equation}
where $nl$ is the number of Lagrangian points closer than the activation distance $d/4$ to the $i$-th point on each filament.
To avoid contacts and overlap between filaments and with the walls, a repulsive force is also implemented. This has the form of a \textit{Morse potential} \citep{liu2004coupling}
\begin{equation}\label{morse}
\phi=D_e\left[\ e^{-2a(r_f-r_e)}-2e^{-a(r_f-r_e)} \right],
\end{equation}
where $D_e$ is the interaction strength, $a$ a geometrical scaling factor, $r_f$ the distance between two elements on two different filaments (or a point on a filament and the wall), and $r_e$ the zero cut-off force distance. The repulsive force between the elements $i$ and $j$ is the derivative of the potential function,
\begin{equation}\label{repulsive_der}
\bm{F}^c_{ij}=-\pi d \frac{d\phi}{dr}\bm{d}_{ij},
\end{equation}
where the factor $\pi d$ is used to convert force per unit area to force per unit length and $\bm{d}_{ij}$ is the unit vector in the direction joining the contact points. Finally, the total repulsive force on the $i$-th element on each filament is obtained as:
\begin{equation}\label{repulsive}
\bm{F}^c_{i}=\sum_{j \neq i}^{nc} \bm{F}^c_{ij},
\end{equation}
where $nc$ is the number of Lagrangian points closer than the cut-off distance $r_e$ to the $i$-th point. Note that, we are neglecting the interaction of filaments with themselves, as we will consider moderate values of flexibility. Furthermore, we found that the results are insensitive to the strength of the repulsive force.
In the case of non-zero surface roughness, a non-negligible friction force may also act on the filaments. An estimate of its magnitude was proposed by \cite{lindstrom2007simulation} as
\begin{equation}\label{friction}
\left|\bm{F}^\mu_{ij}\right|=\mu_f \left|\bm{F}^c_{i}\right|,
\end{equation}
where $\mu_f$ is the friction coefficient. In present study, since we focus on inertial and elastic effects at relatively low volume fractions, we neglect the frictional effects.
\subsubsection{Solution of the filament equations}
\begin{figure}
     \centering
    \includegraphics[width = .8\textwidth,trim={0 18cm 0 0},clip]{{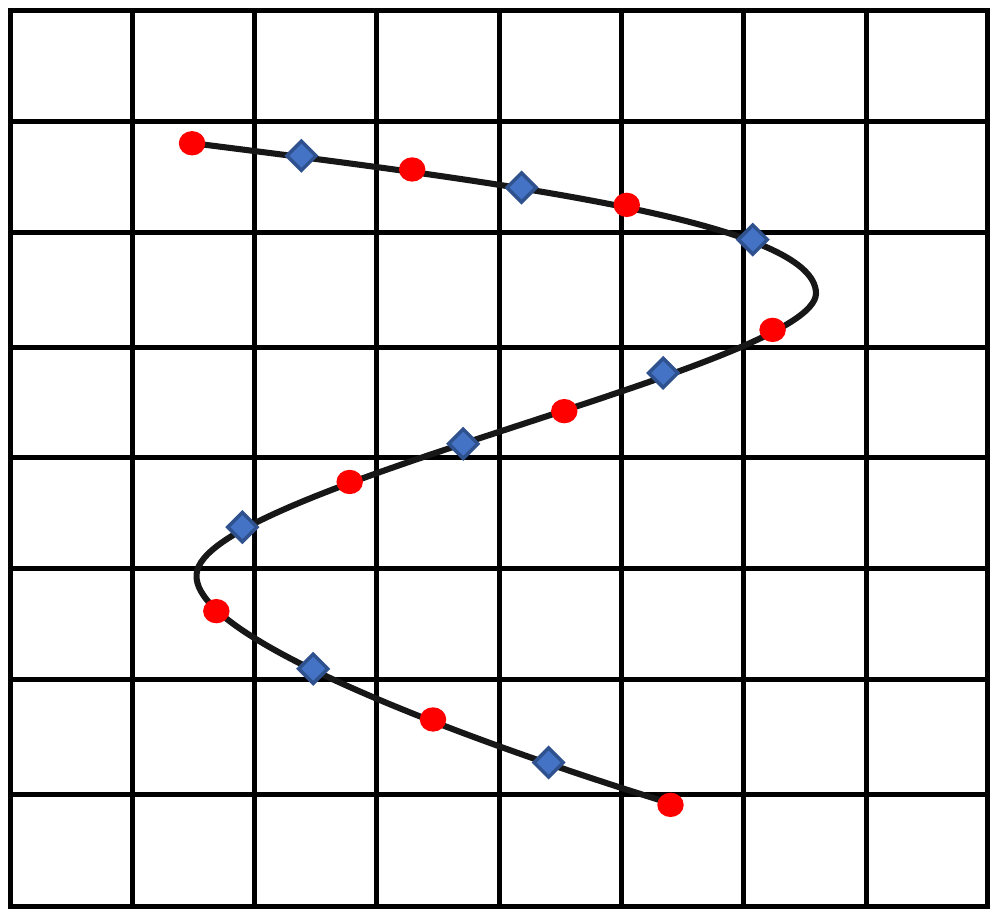}}
     \caption{Schematic of the Eulerian and the staggered Lagrangian grids. The red circles denote the Lagrangian points which the position of the filaments is defined and the blue diamonds show the Lagrangian points used to compute tension.}
     \label{fibre_schematic}
 \end{figure}

In order to solve the filament equations (\ref{Eul_Ber}) and (\ref{inext}), we follow the two-step method proposed by \cite{huang2007simulation} for inertial filaments. Note that, in the case of neutrally buoyant filaments, i.e. $\left( \rho_f/A_f-\rho\right)=0$, the LHS of equation (\ref{Eul_Ber}) vanishes and in order to avoid the singularity of the coefficient matrix, the discretized equation (\ref{Eul_Ber}) is modified as in \cite{pinelli2016pelskin}
\begin{equation} \label{Eul_Ber_mod}
\frac{\partial^2 \bm{X}}{\partial t^2} =\frac{\partial^2 \bm{X_f}}{\partial t^2} +  \frac{\partial}{\partial s} \left( T \frac{\partial \bm{X}}{\partial s}\right) - B \frac{\partial^4 \bm{X}}{\partial s^4}- \bm{F} + \bm{F}^c,
\end{equation}
where the LHS is the filament acceleration, whereas the first term on the RHS is the fluid particle acceleration. By doing so, it is then feasible to proceed as detailed by \cite{huang2007simulation}. In particular, \textit{i)} we solve a Poisson equation for the tension, derived by combining equations (\ref{Eul_Ber_mod}) and (\ref{inext}),
\begin{equation} \label{poi_tension}
\frac{\partial \bm{X}}{\partial s} \cdot \frac{\partial^2}{\partial s^2} \left( T \frac{\partial \bm{X}}{\partial s}\right)= \frac{1}{2}\frac{\partial^2}{\partial t^2} \left(\frac{\partial \bm{X}}{\partial s} \cdot \frac{\partial \bm{X}}{\partial s}\right)- \frac{\partial^2\bm{X}}{\partial t\partial s}\cdot\frac{\partial^2\bm{X}}{\partial t\partial s}- \frac{\partial \bm{X}}{\partial s}\cdot\frac{\partial}{\partial s} \left(\bm{F}^a+\bm{F}^b + \bm{F}^c - \bm{F}\right), 
\end{equation}
where $\bm{F}^a=\frac{\partial^2 \bm{X_f}}{\partial t^2}$ is the acceleration of the fluid particle at the filament location and $\bm{F}^b=-B \frac{\partial^4 \bm{X}}{\partial s^4}$ the bending force; \textit{ii)}  we solve equation (\ref{Eul_Ber_mod}) to obtain the updated position of the Lagrangian points defining the filament. 
The equations introduced above in the continuum notation are discretised by a second-order finite-difference scheme on a staggered grid for tension and position \citep{huang2007simulation}. A schematic of the grid points is presented in figure \ref{fibre_schematic}. We solve the Poisson equation (\ref{poi_tension}) for the tension using a predicted position $\bm{X}^{*}=2\bm{X}^{n}-\bm{X}^{n-1}$, where $\bm{X}^{n}$ and $\bm{X}^{n-1}$ are the solutions at previous times; to find the new position of the filaments at time $t_{n+1}$, the updated value of the tension $T$ is used in equation (\ref{Eul_Ber_mod})  whose discrete version is
 \begin{equation} \label{discrete_Eul}
 \frac{\bm{X}_i^{n+1}-2\bm{X}_i^{n}+\bm{X}_i^{n-1}}{\Delta t^2}=\frac{\bm{X}_i^{n}-2\bm{X}_i^{n-1}+\bm{X}_i^{n-2}}{\Delta t^2}+S,
 \end{equation}
where $S$ is a source term including the discrete form of the tension, bending and forcing terms. Equation (\ref{discrete_Eul}) reduces to a penta-diagonal matrix which is inverted by Gaussian elimination to obtain $\bm{X}^{n+1}$.

 To correctly obtain the filament rotation within our one-dimensional model, we consider four ghost points at a small radial distance ($\approx \Delta x =\frac{d}{2}$) around each of the Lagrangian points used to compute the hydrodynamic forces. These 4 points are then used to evaluate the moment exerted by the fluid on the filament
\begin{equation} \label{shear_moment}
\bm{M}=\bm{r}\times\bm{F},
\end{equation}
where $\bm r$ is the position vector connecting the main Lagrangian points to the ghost points. Note that, the moment at the two ends of the filaments should be set to $\bm{M}=-\bm{r}\times\bm{F}$ in order to satisfy the zero moment condition. The effect of the shear moment is then introduced in the Euler-Bernoulli equation (\ref{Eul_Ber_mod}) to provide the correct rotation
\begin{equation} \label{Eul_Ber_mod2}
\frac{\partial^2 \bm{X}}{\partial t^2} =\frac{\partial^2 \bm{X_f}}{\partial t^2} +  \frac{\partial}{\partial s} \left( T \frac{\partial \bm{X}}{\partial s}\right) - B \frac{\partial^4 \bm{X}}{\partial s^4}-\frac{\partial}{\partial s}D\left(\bm{M}\right)- \bm{F} + \bm{F}^f,
\end{equation}
where $D$ is defined as
\begin{equation} \label{operator}
D\left(M_i\right)=\sum_{i\neq j}M_j.
\end{equation}

Note that the contribution from the shear moment appears in the Poisson equation for the tension which is solved in combination with the Euler Bernoulli equations. The fluid equations are solved with a second-order finite-difference method on a fix staggered grid. The equations are advanced in time by a semi-implicit fractional step-method, where the second order Adams-Bashforth method is used for the convective terms, a Helmholtz equation is built with the diffusive and temporal terms, and all other terms are treated explicitly \citep{alizad2017numerical}.

To summarise the numerical algorithm, the following procedure is performed at each time step:
\begin{enumerate}
    \item the fluid velocity is interpolated on the Lagrangian points using Eq. (\ref{interpolation});
    \item the Lagrangian force is computed from Eq. (\ref{force});
    \item the Lagrangian force is spread onto the Eulerian grid by Eq. (\ref{spread});
    \item the lubrication and/or repulsive force is computed using Eqs.\ (\ref{lubcorri}) and (\ref{repulsive_der});
    \item the predicted value of the position $\bm{X}^{*}$ is obtained: $\bm{X}^{*}=2\bm{X}^{n}-\bm{X}^{n-1}$;
    \item the tension is computed by solving Eq. (\ref{poi_tension});
    \item the new filament  position is obtained by solving Eq. (\ref{discrete_Eul});
    \item the fluid equations are advanced in time.
\end{enumerate}

\subsection{Code validation}
\subsubsection{A hanging flexible filament under gravity}
\begin{figure} 
\centering
\vspace{0.5cm}
\includegraphics[width=0.4\textwidth]{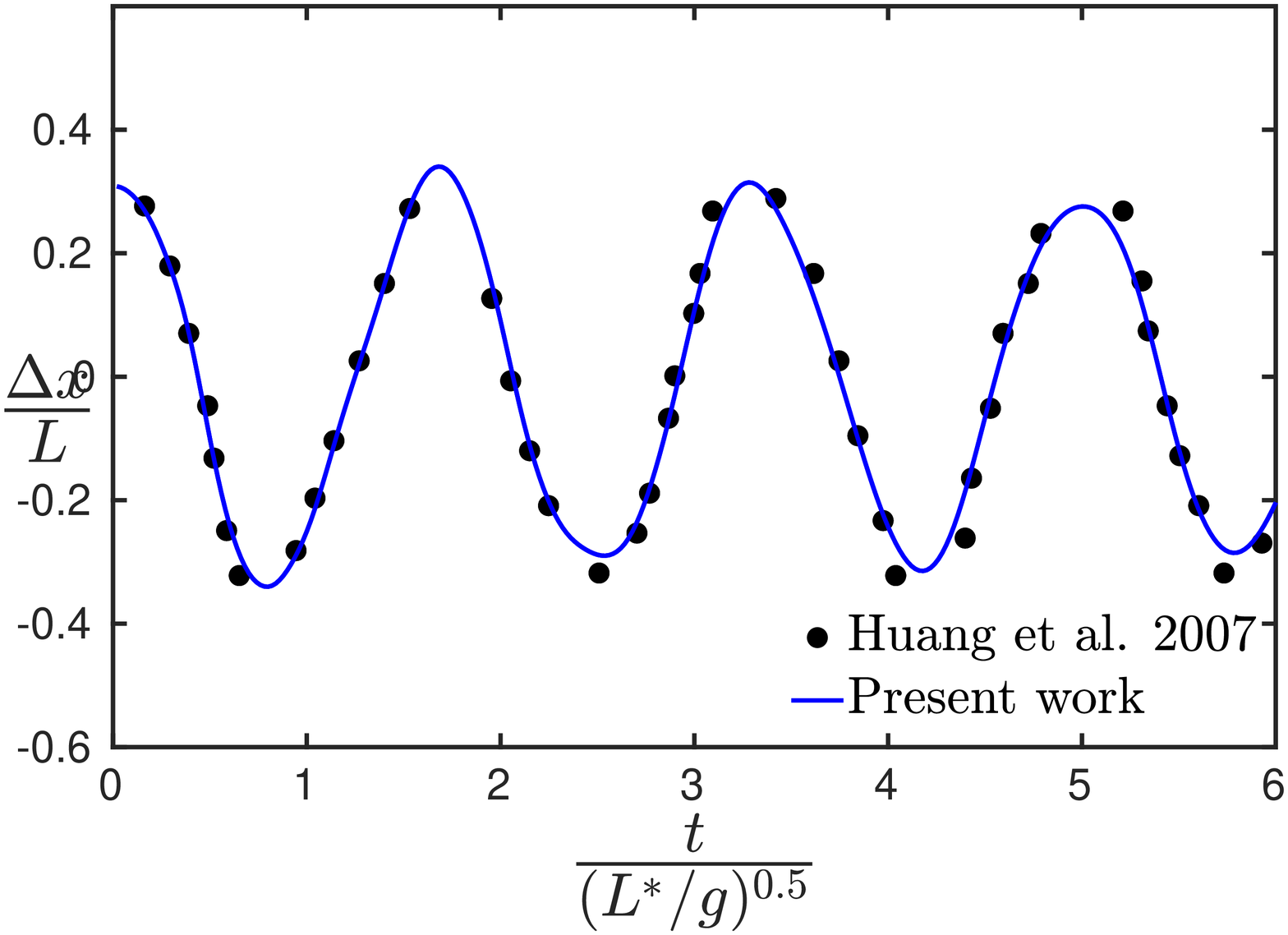}
\hspace{0.75cm}
\includegraphics[width=0.4\textwidth, angle =0]{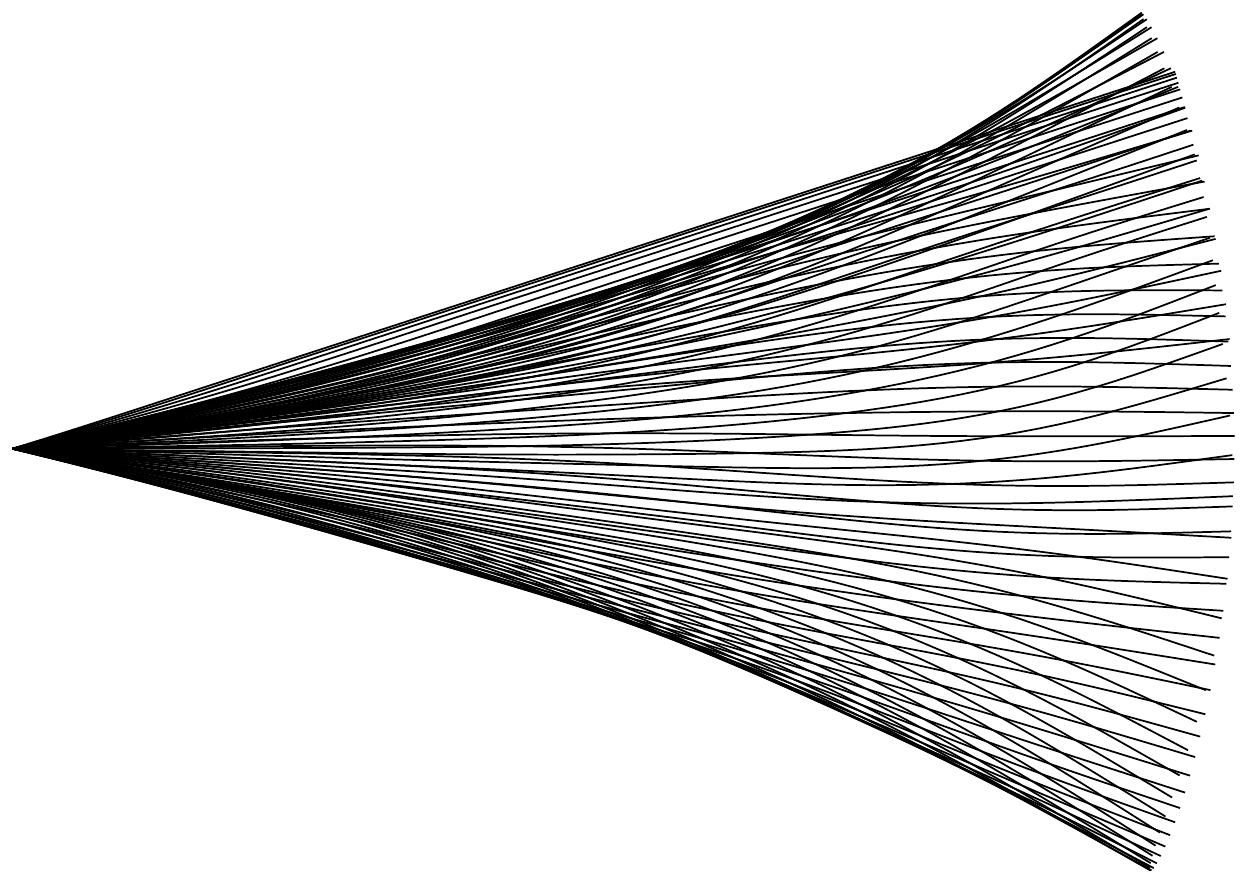}
\caption{Oscillations of a single hanged filament under gravity without flow, visualization on the right, for $\frac{B^\ast}{\rho_f g L^3}=0.01$. Our results, shown with the solid line, are compared with those reported by \cite{huang2007simulation} (solid dots).}
\label{huang}
\end{figure}
In order to validate the implementation of the structural solver, we study the oscillations of a hanging hinged filament under gravity, in the absence of any external flow, as done by \cite{huang2007simulation}. In this case, the filament oscillates at its natural frequency (first mode). The results obtained with a resolution of 30 Lagrangian points per filament length are displayed in figure~\ref{huang}, where we find a good agreement with the numerical results from the literature. We also successfully tested the oscillation frequencies of natural frequencies of the higher order modes  against the analytical solution (not reported here).

\subsubsection{A single rigid filament in a shear flow}
\begin{figure}
\centering
\includegraphics[width=0.48\textwidth]{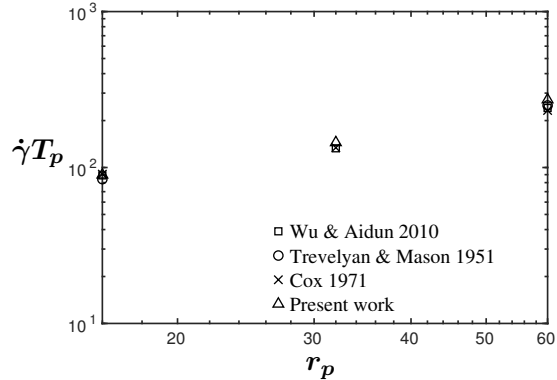}
\put(-95,-10){\large $\bm{r_p}$}
\put(-206,70){\large $\bm{\dot{\gamma}T_p}$}
\caption{Rotation periods $T_p$ of rigid filaments in shear flows with different aspect ratios $r_p$:  $\triangle$ indicate our simulations, compared with the results by \cite{trevelyan1951particle} denoted by $\bigcirc$, \cite{cox1971motion} represented with $\times$ and \cite{wu2010method}, $\square$.}
\label{rotation}
\end{figure}
We now consider a single rigid filament in a shear flow and compare its period of rotation with those analytically derived by \cite{cox1971motion}, computed numerically by \cite{wu2010method} and measured experimentally by \cite{trevelyan1951particle}. The rigid filament behaviour is simulated by setting $B=150$ which ensures a negligible bending of the filaments. As previously mentioned, in our discretization approach we consider four additional ghost points around each Lagrangian points at a small radial distance ($\approx \Delta x =\frac{d}{2}$). These 4 points are used to evaluate the correct moment exerted by the fluid on the filament, which is then introduced in the Euler-Bernoulli equation  (\ref{Eul_Ber_mod}) as forces normal to the filament axis. Figure \ref{rotation} compares the period of rotation as a function of the filament aspect ratio $r_p$ obtained from our simulations with the results in the literature mentioned above, and a very good agreement is found. The simulations are performed in a computational box of size $5L^* \times 8L^* \times 5L^*$, where $L^*$ is the filament length, discretised with $80 \times 128 \times 80$ grid points, respectively. 17 Lagrangian points are used for the filament.

\subsubsection{Deformable filaments in oscillatory flow}
\begin{figure} 
\centering
\includegraphics[width=0.3\textwidth]{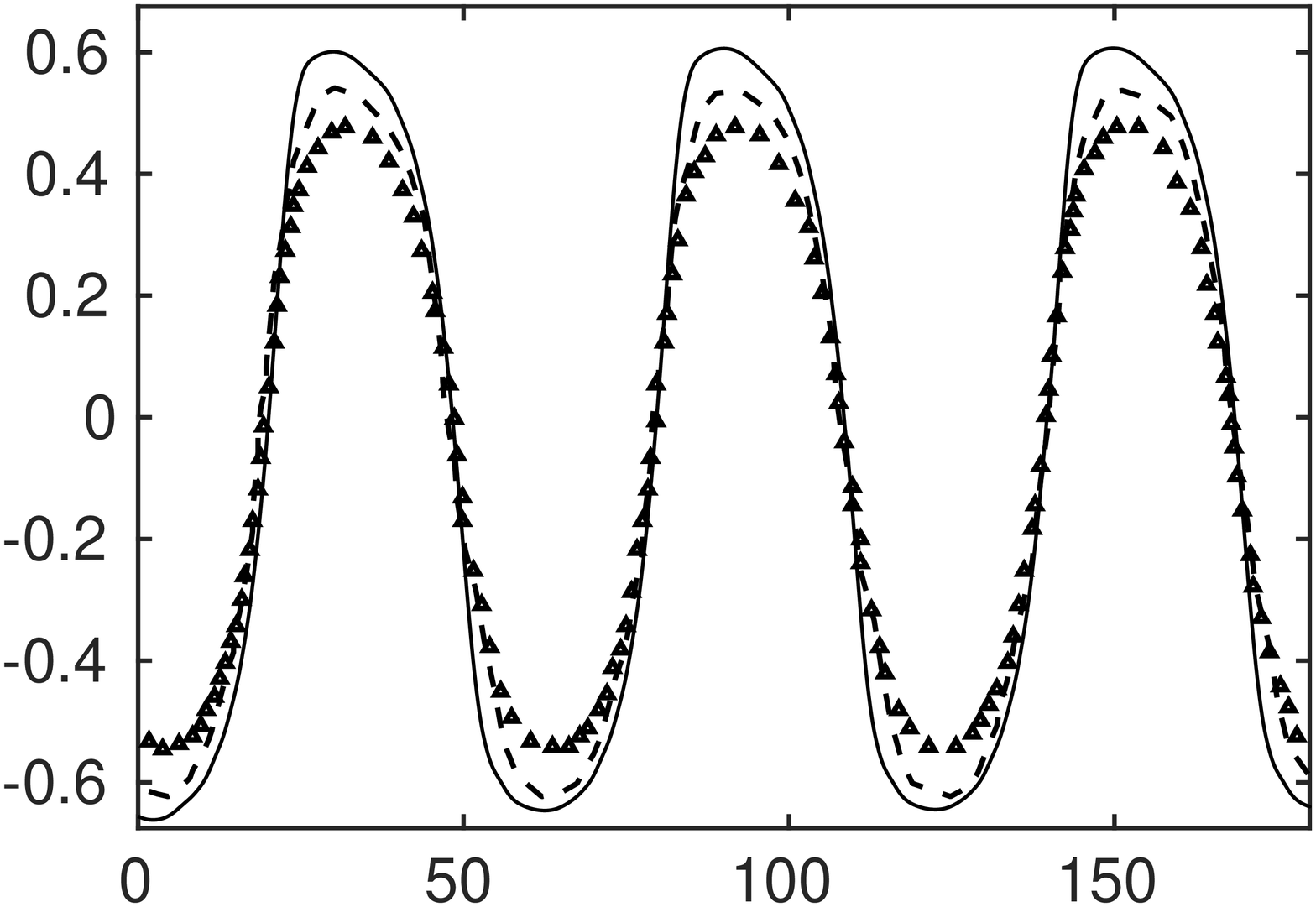}
\hspace{.5cm}
\includegraphics[width=0.55\textwidth]{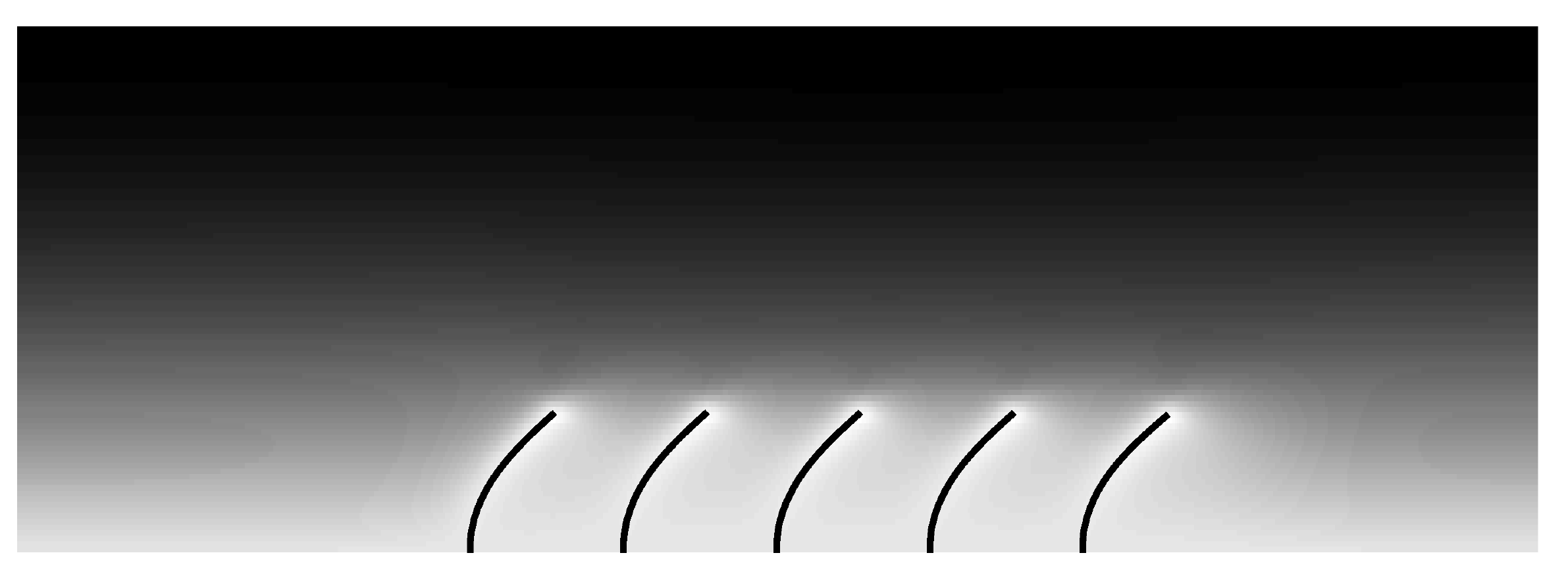}
\put(-299,-15){\large $\frac{t U_{max}}{L^\ast}$}
\put(-360,48){\large $\frac{\Delta x}{L^*}$}
\put(-360,75){\large $(a)$}
\put(-225,75){\large $(b)$}
\caption{a) Time history of the displacement of the filament free-end in an oscillatory flow (solid line) compared to the experimental (dashed line) and numerical (triangles) results by \cite{pinelli2016pelskin}. b) Instantaneous visualisation of the row of filaments in the half channel. The background colors indicate the streamwise velocity.}
\label{osc}
\end{figure}
Finally, we consider a row of five flexible filaments clamped at the bottom wall of a channel with a flow driven by an oscillatory pressure gradient and compare our results with those obtained from the experiments and computations presented in \cite{pinelli2016pelskin}. In this simulation, the bulk Reynolds number based on maximum bulk velocity $U_{bulk}$ is equal to $Re=40$, the bending stiffness is $\frac{B^\ast}{\rho_f U^2_{bulk}{L^\ast}^2}=3.81$, and the oscillation frequency is $\frac{1}{60} \frac{U_{bulk}}{L^*}$. 
The filament length is $\frac{1}{6}$ of the channel height and the separation distance between the filaments is equal to one filament length. The size of the computational domain is $10L^* \times 6L^* \times 5L^*$ in the streamwise, wall-normal and spanwise directions, discretised by $160 \times 96 \times 80$ grid points; $17$ Lagrangian points are used to describe a single filament. Figure \ref{osc} shows the oscillations of the right-most filament of the row obtained from our simulations. Both the frequency and the amplitude of the oscillations match the computational and experimental data reported in \cite{pinelli2016pelskin}. We observe that the difference between the two numerical approaches is small; in particular, the slightly different amplitude may be due to the different numerical approaches and the uncertainty of the experiments.

\section{Flow configuration and suspension rheology}
\begin{figure} 
\centering
\includegraphics[width=\textwidth]{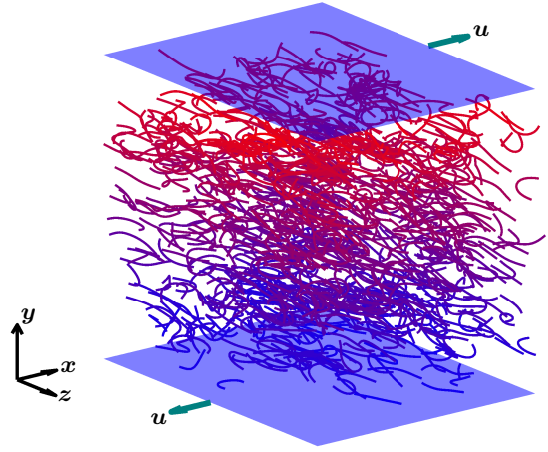}
\put(-110,192){$\bm{u}$}
\put(-232,45){$\bm{u}$}
\put(-266,65){$\bm{x}$}
\put(-281,85){$\bm{y}$}
\put(-267,54){$\bm{z}$}
\caption{Schematic of the configuration and reference frame adopted in this study. The visualisation refers to a suspension of flexible filaments at $Re=10$ with  volume fraction $\phi=0.018$. }
\label{geo}
\end{figure}
We study suspensions of flexible filaments at moderate volume fractions in a Couette flow, focusing on the role of inertia and flexibility on the suspension rheology. We consider stiff and flexible filaments suspended in a channel with the upper and lower walls moving with opposite velocities in the streamwise $x$-direction; no-slip and no-penetration conditions are enforced on the moving walls, while periodicity is assumed in the homogeneous streamwise and spanwise directions. Initially, the filaments are randomly distributed in the channel. Figure~\ref{geo} depicts the flow configuration and the coordinate system used in the present study, where the computational domain has size $5L^* \times 5L^* \times 8L^*$.The filament aspect ratio is set to $r_p=1/16$ for all cases. In the present study, quantities are made dimensionless by the viscous scales, thus the non-dimensional bending stiffness is defined as 
\begin{equation} \label{gammad}
\tilde{B}=\frac{B^\ast}{\mu \dot{\gamma}L^4},
\end{equation}
and is related to the bending stiffness previously reported in equation (\ref{Eul_Ber}) by
\begin{equation} \label{gammadgamma}
\tilde{B}=\frac{\pi}{4}r_p^2 Re \, B.
\end{equation}
The solid volume fraction of the suspension is
\begin{equation} \label{volf}
\phi=  \frac{n\pi r_p^2}{4 V},
\end{equation}
where $n$ is the number of filaments in the computational box and $V$ the volume of the computational domain.

We perform a parametric study to assess the role of inertia, flexibility and volume fraction on the suspension flow. In particular, we vary the Reynolds number in the range $0.1\leq Re \leq10$, the bending rigidity in the range $0.005\leq \tilde{B} \leq 0.5$ and the volume fractions in the range $0.0265 \leq \phi \leq 0.106$; table \ref{table:num} reports all the cases considered in the present work. In total, 24 configurations have been considered. For the filaments considered here, the so-called concentrated regime \citep{wu2010numerical}, when the filament-filament interactions become the dominant in determining the macroscopic suspension behavior, is reached for volume fractions $\phi \geq 0.053$ ($\frac{n {L^\ast}^3}{V}>\frac{1}{r_p}$); in this case, suitable models for lubrication and contact forces are necessary. 

In all our simulations, we use $80\times128\times80$ grid points in the streamwise, wall normal and spanwise directions to discretise the computational domain, while $17$ Lagrangian points are used to describe each suspended filament, where the resolution has been chosen to properly resolve the cases with most flexible filaments. The time step necessary to properly capture the full filament dynamics is of the order of $\Delta t \approx 10^{-5}$; note that, the main time-step constraint is determined by the elastic and lubrication forces in all the cases. We performed additional simulations with domain size, space and time resolution increased by a factor of $2$ for the most demanding cases and found that the difference in the suspension viscosity is lower than $2\%$.
\begin{table}
\begin{center}
\begin{tabular}{c c c c c c}
\multicolumn{2}{c}{Eulerian grid points} & \multicolumn{4}{c}{$80\times 128 \times 80$} \\
\hline
\multicolumn{2}{c}{Lagrangian grid points} & \multicolumn{4}{c}{$17$} \\
\hline
\multicolumn{2}{c}{$Re$} & \multicolumn{4}{c}{$\tilde{B}$} \\
\hline
& & 0.005 & 0.02 & 0.05 & 0.5 \\ 
 & 0.1 & $\phi=0.053$ & \vtop{\hbox{\strut $\phi=0.0265$}\hbox{\strut $\phi=0.053$}\hbox{\strut $\phi=0.0795$}\hbox{\strut $\phi=0.106$}} & $\phi=0.053$ & \vtop{\hbox{\strut $\phi=0.0265$}\hbox{\strut $\phi=0.053$}\hbox{\strut $\phi=0.0795$}\hbox{\strut $\phi=0.106$}}  \\
\hline
 & 1 &  $\phi=0.053$ & $\phi=0.053$ & $\phi=0.053$ & $\phi=0.053$\\ 
\hline
 & 10 & $\phi=0.053$ & \vtop{\hbox{\strut $\phi=0.0265$}\hbox{\strut $\phi=0.053$}\hbox{\strut $\phi=0.0795$}\hbox{\strut $\phi=0.106$}} & $\phi=0.053$ & \vtop{\hbox{\strut $\phi=0.0265$}\hbox{\strut $\phi=0.053$}\hbox{\strut $\phi=0.0795$}\hbox{\strut $\phi=0.106$}} 
\end{tabular}
\caption{Summary of the numerical simulations performed in this study. The table reports the Reynolds number $Re$, the volume fraction $\phi$ and the bending rigidity $\tilde{B}$ of all cases presented in this study.}
\label{table:num}
\end{center}
\end{table}

\subsection{Rheology of filament suspensions}
The rheological behavior of the suspensions is presented in terms of the relative viscosity
\begin{equation}
\eta = \frac{\mu_{eff}}{\mu},
\end{equation}
where $\mu$ is the viscosity of the carrier fluid and $\mu_{eff}$ is the effective viscosity of the suspension. The relative viscosity can be rewritten in terms of the bulk shear stress as
\begin{equation} \label{mu_eff}
\eta =1+\bar{\Sigma}^f_{xy},
\end{equation}
where  $\bar{\Sigma}^f_{xy}$ is the time and space average of the shear stress arising from the presence of the filaments, non-dimensionalised by the imposed shear rate $\dot{\gamma}_{xy}$ and the viscosity $\mu$. The normal stress differences are used to describe the Non-Newtonian behaviour of the suspension, and are defined as 
\begin{equation}
N_1= \bar{\Sigma}_{xx}-\bar{\Sigma}_{yy}, \quad  N_2= \bar{\Sigma}_{yy}-\bar{\Sigma}_{zz}.
\end{equation}

To compute the total stress in the suspension and to differentiate all the different contributions, we follow the derivation first proposed by \cite{batchelor1970stress} for a suspension of rigid spherical particles and adapt it to the case of flexible filaments, see also \cite{batchelor1971stress,wu2010method}. The dimensionless total stress reads
\begin{equation} \label{batchelor}
\Sigma_{ij} =Re\left[\frac{1}{V}\int_{V-\Sigma V_0}\left(-P \delta_{ij}+ \frac{2}{Re}e_{ij}\right) dV+\frac{1}{V} \sum \int_{ V_0}\sigma_{ij}, dV-\frac{1}{V}\int_{V} u'_i u'_j dV\right],
\end{equation}
where $V$ is the total volume under investigation and $V_0$ the volume of each filament; $e_{ij}=\frac{\partial {u_i}}{\partial {x_j}}+\frac{\partial {u_j}}{\partial {x_i}}$ represents the strain rate tensor and ${\bm{u}}' $ the velocity fluctuations. The first term on the right hand side of equation (\ref{batchelor}) represents the fluid bulk viscous stress tensor, the second term the stress generated by the fluid-solid interaction forces and the last term the stress generated by the velocity fluctuations in the fluid (the Reynolds stress tensor). We may write the total stress as the summation of the fluid and filament stress tensors:
\begin{equation} \label{viscfil}
\Sigma_{ij}=\bar{\Sigma}^0_{ij}+\bar{\Sigma}^f_{ij},
\end{equation}
where 
\begin{equation} \label{sigma0}
\begin{aligned}
        & \bar{\Sigma}^0_{ij}=\frac{Re}{V}\int_{V-\Sigma V_0}\left(-P \delta_{ij}+ \frac{2}{Re}e_{ij}\right) dV,\\
        & \bar{\Sigma}^f_{ij}=\frac{Re}{V} \sum \int_{ V_0}\sigma_{ij} dV-\frac{Re}{V}\int_{V} u'_i u'_j dV.
        \end{aligned}
\end{equation}
The fluid-solid interaction stress can be decomposed into two parts \citep{batchelor1970stress}:
\begin{equation} \label{stressdecomp}
\int_{V_0}\sigma_{ij} dV=\int_{A_0}\sigma_{ik}x_jn_k dA-\int_{V_0}\frac{\partial \sigma_{ik}}{\partial x_k}x_j dV,
\end{equation}
where $A_0$ represents the surface area of each filament. The first term is called the stresslet and the second term indicates the acceleration stress \citep{guazzelli2011physical}. For neutrally buoyant filaments, when the relative acceleration of fluid and the filament is zero, the second term in (\ref{stressdecomp}) is identically zero. $\sigma_{ik}n_k$ is the force per unit area acting on the filaments \citep{batchelor1971stress}, that for slender bodies can be rewritten as
\begin{equation} \label{slender}
\int_{A_0}\sigma_{ik}x_jn_k dA=-r_p^2\int_{L}\bm{F}_{i}x_j ds,
\end{equation}
where the term $r_p^2$ arises from choosing the linear density instead of the volume density as scale for the fluid-solid interaction force. Finally, the filament stress is
\begin{equation} \label{batchelor1}
\Sigma^f_{ij}=-\frac{Re{r_p}^2}{V} \sum \int_L\bm{F}_{i}x_j ds-\frac{Re}{V}\int_{V} u'_i u'_j dV.
\end{equation}
From the results of our simulations, we observe that the last term, related to the velocity fluctuations, is very small compared to the stresslet and can be thus neglected for the range of Reynolds numbers considered here. This is consistent with the behavior of rigid particles for the same Reynolds numbers, $\mathcal{O}(10)$, as shown in \cite{alghalibi2018interface}.

\section{Results}
\subsection{Suspensions of rigid fibers in shear flow}
\begin{figure} 
\centering
\includegraphics[width=0.7\textwidth]{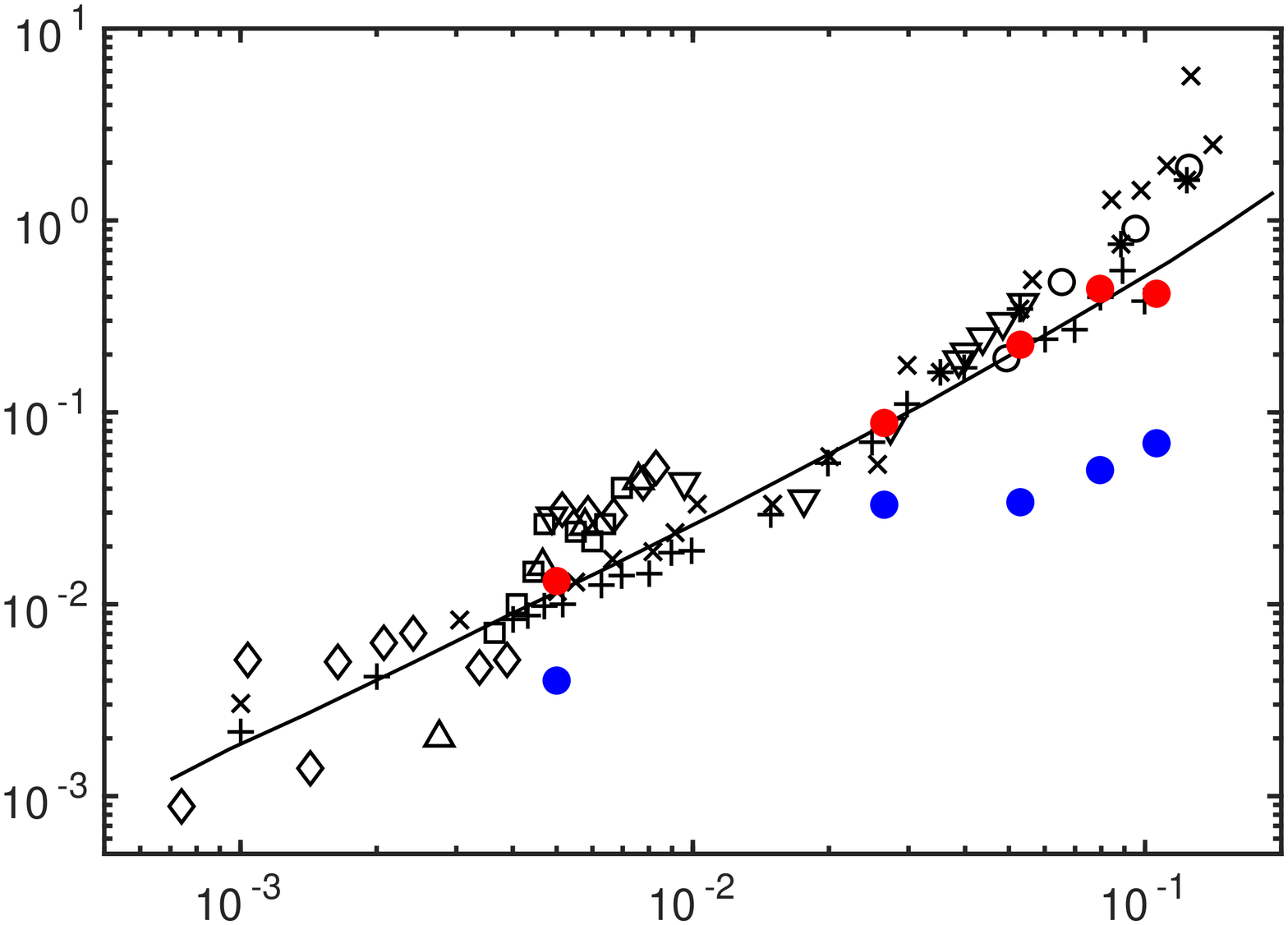}
\put(-310,110){\Large $\eta-1$}
\put(-130,-20){\Large $\phi$}
\caption{Relative viscosity versus solid volume fraction for suspensions of rigid fibers. The red and blue filled symbols refer to the present numerical results with and without the lubrication correction. The open symbols denote the experimental data of \cite{blakeney1966viscosity}, while the filled black symbols the experimental data of \cite{bibbo1987rheology}. The $+$ symbols show the numerical results by \cite{lindstrom2008simulation}, and $\times$ 
those by \cite{lindstrom2008simulation}, including contact friction between filaments. The solid line represents the theoretical prediction of \cite{liu2004coupling} for a suspension of fibers rotating only in the shear plane.}
\label{rigid}
\end{figure}

We start our analysis by comparing the relative viscosity of concentrated rigid fiber suspensions at negligible flow inertia obtained from our simulations with the theoretical, numerical and experimental results reported in the literature. In particular, we discuss here the role of the short-range friction model for fibers. Our results are obtained for a Reynolds number equal to $Re=0.1$ and with flexibility $\tilde{B}=0.5$, which properly reproduces rigid filaments. The data are presented in figure \ref{rigid} showing that results from the present numerical simulations are within the range predicted by the numerical and experimental results in the literature, as well as  the theoretical prediction of \cite{liu2004coupling} for a suspension of fibers rotating only in the shear plane. In order to show the importance of the lubrication correction, we also display results obtained without it: in this case the suspension viscosity is strongly under-predicted, especially at high volume fractions, resulting in large differences between our results and the experimental data. The test confirms that within the framework of our numerical scheme and with the chosen grid resolution, the short-range interactions are indeed important when considering concentrated regimes.

\subsection{Suspensions of flexible filaments at fixed volume fraction}
\begin{figure} 
\centering
\includegraphics[width=0.42\textwidth]{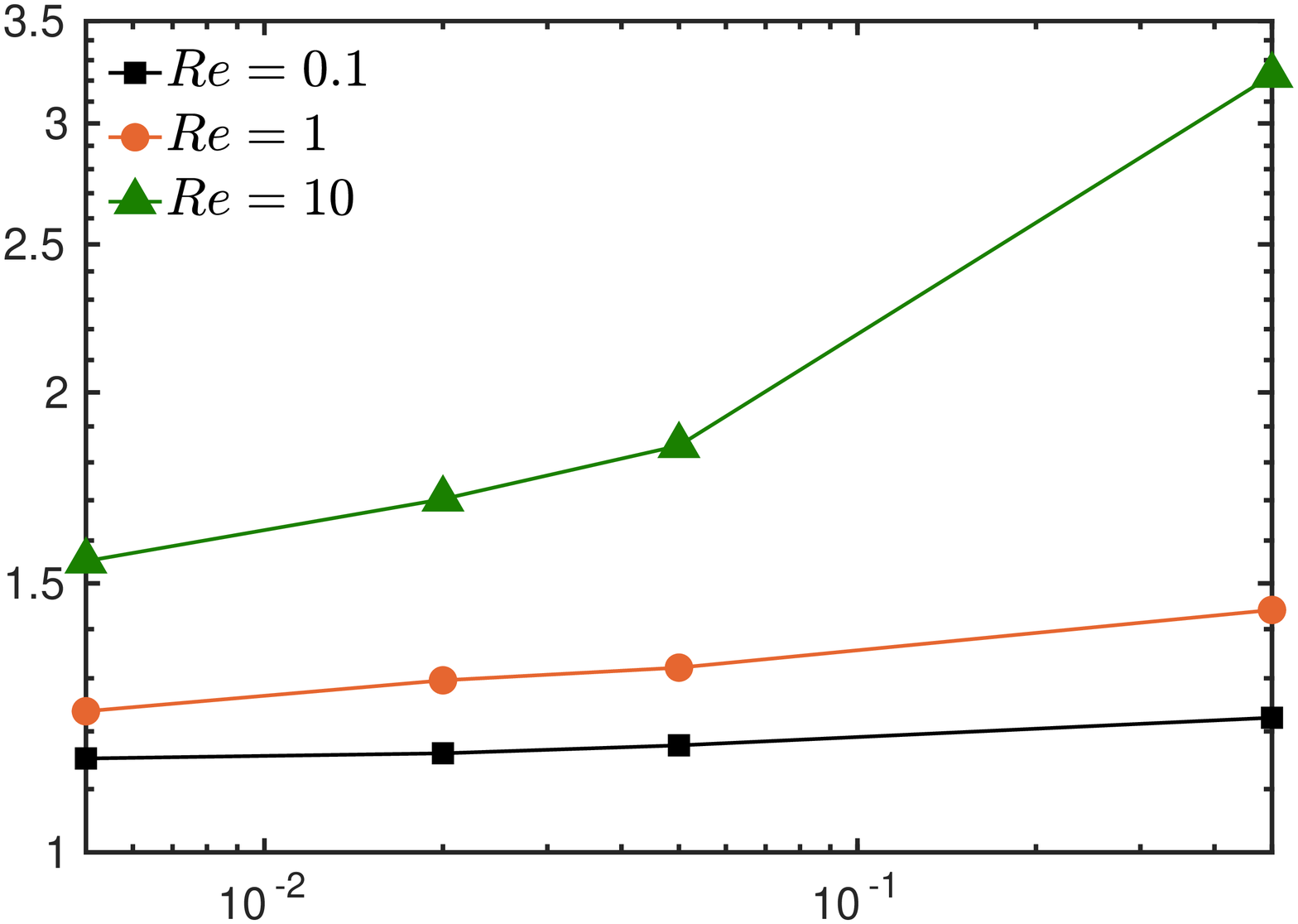} 
\put(-85,-10){\large $\tilde{B}$}
\put(-175,70){\large $\eta$}
\put(-175,110){\large $a)$}
\hspace{1cm}
\includegraphics[width=0.42\textwidth]{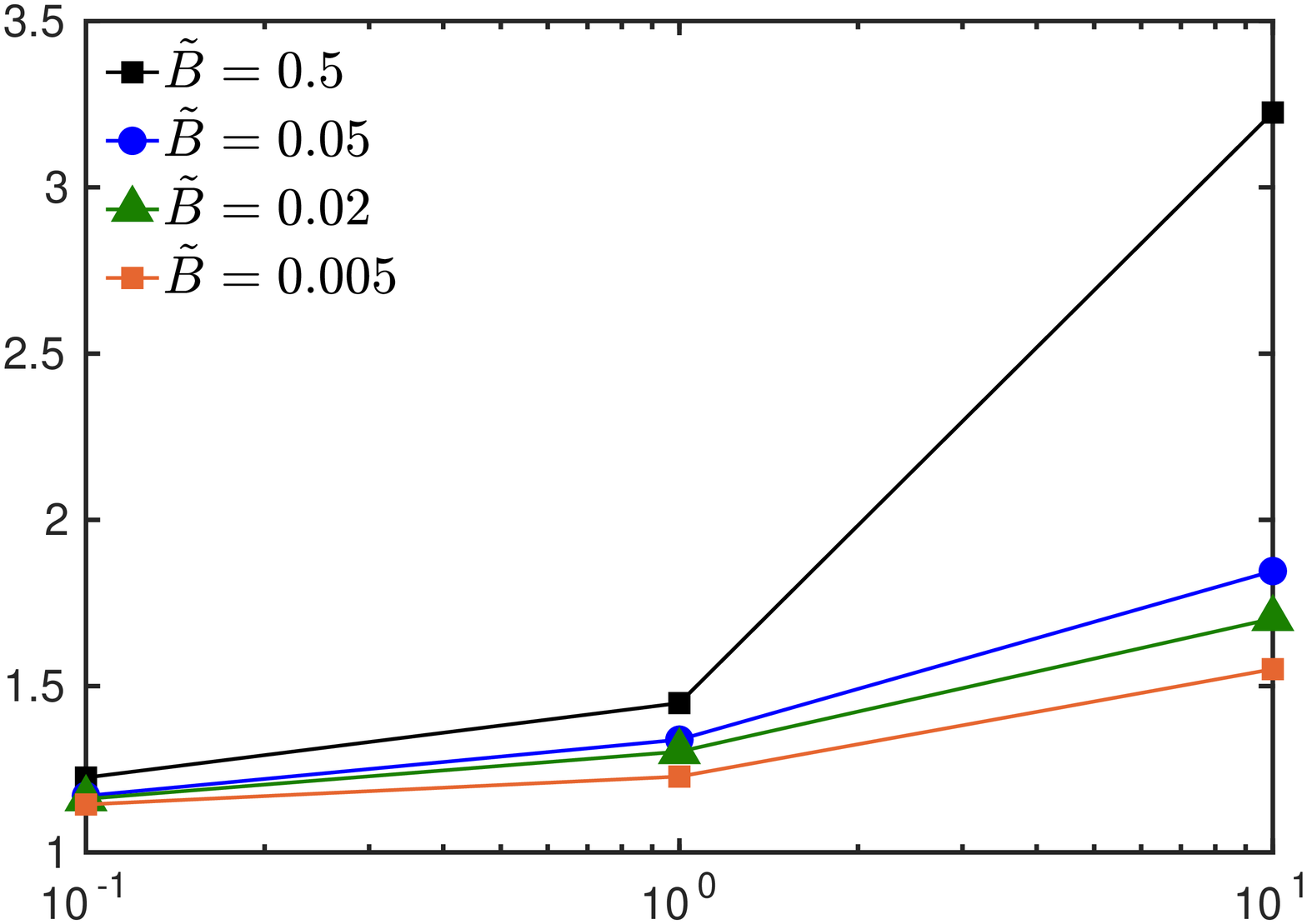}
\put(-175,70){\large $\eta$}
\put(-175,110){\large $b)$}
\put(-90,-10){\large $Re$}
\caption{Relative viscosity of filament suspensions versus (a) the filament bending rigidity and (b) the Reynolds number. The suspension volume fraction is fixed to $\phi=0.053$ for all cases.}
\label{fig_visc}
\end{figure}
We now analyse the suspension behavior at fixed volume fraction, $\phi=0.053$, and vary the Reynolds number and filament flexibility. First, we report in figure \ref{fig_visc} the dependence of the suspension relative shear viscosity on the bending rigidity (left panel) and the Reynolds number (right panel). Figure \ref{fig_visc}a shows that the viscosity increases with the bending rigidity, i.e.,~the viscosity decreases as the filaments are more flexible. This result is in agreement with the simulations by \cite{switzer2003rheology,sepehr2004rheological} for flexible fibers, and also to the results pertaining the case of deformable particles, capsules and droplets, see e.g.\ \cite{reasor2013rheological,matsunaga2016rheology,rosti2018suspensions,rosti_de-vita_brandt_2019a}. These observations are in contrast with the results of \cite{wu2010numerical} where larger viscosities are obtained for more flexible cases.  
This difference can be explained by the different physical objects under consideration: \citet{wu2010numerical} and \citet{joung2001direct} considered chains of interconnected rigid particles which can twist and bend in their joints, while we consider continuously flexible filaments that can only bend. Suspensions of elastic elongated objects can therefore display different behavior: viscosity decreasing with deformability \citep{switzer2003rheology,sepehr2004rheological} or increasing with it \citep{wu2010numerical,joung2001direct}. Note also that, although \citet{switzer2003rheology} and \citet{sepehr2004rheological} adopt a model similar to the one used by \citet{wu2010numerical}, they observe the same behaviour as in our results which may be explained by the different aspect ratio considered: indeed, the former authors considered fibers whose aspect ratio is at least $5$ times smaller than those studied by \citet{wu2010numerical}. Note also that the relative viscosity changes with the flexibility in a very similar way to shear thinning fluids: with decreasing $\tilde{B}$, the relative viscosity approaches a constant value while the other plateau for high $\tilde{B}$ is not captured in the range of rigidity considered in this work.

Figure \ref{fig_visc}b displays the same data of \ref{fig_visc}a, now as a function of the Reynolds number, in order to highlight how inertia affects the suspension viscosity. We observe that $\eta$ increases with the Reynolds number, especially for the most stiff cases; this indicates that inertial effects are more evident for rigid rods than for flexible filaments. 

\begin{figure} 
\centering
\includegraphics[width=0.4\textwidth]{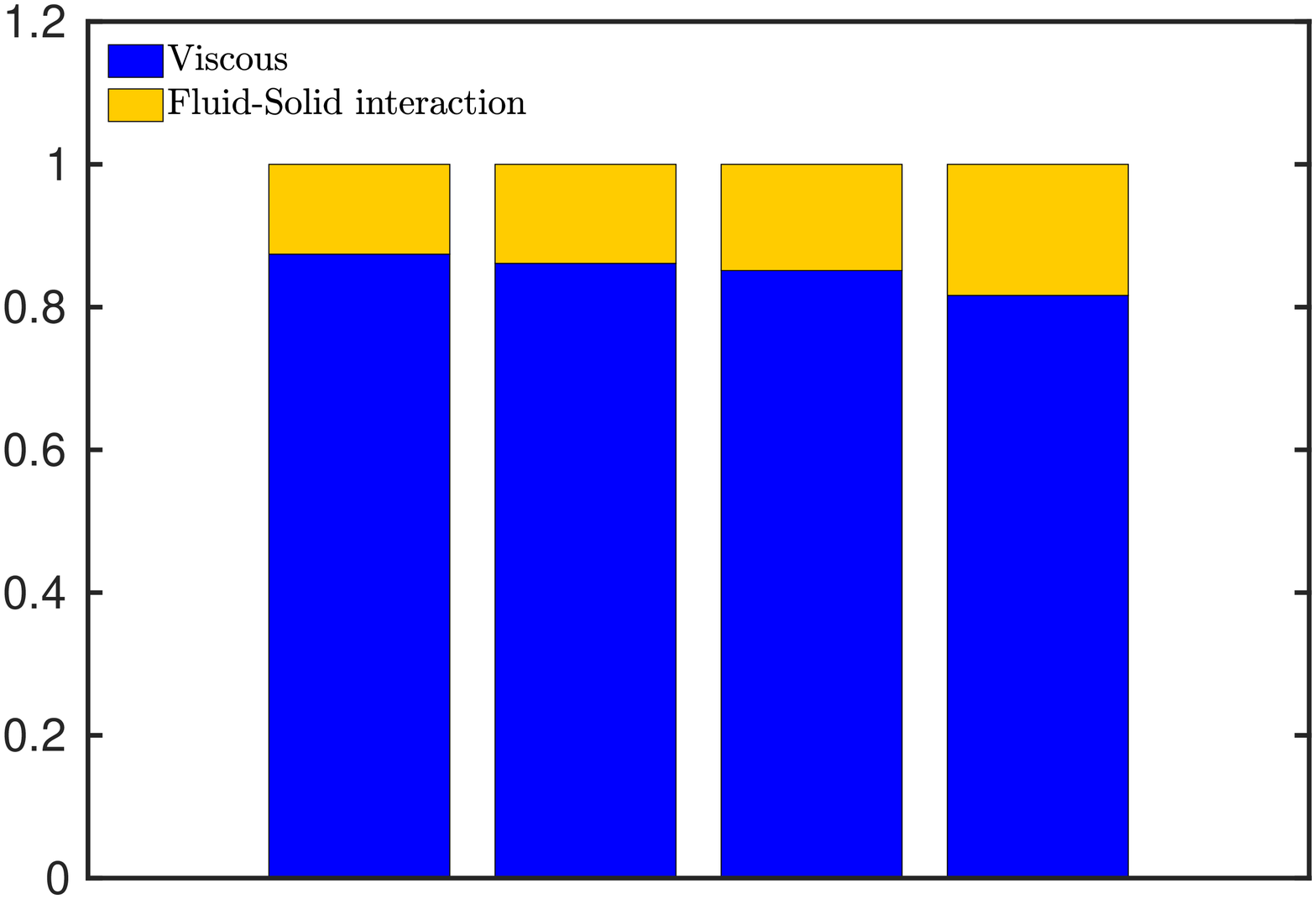} 
\put(-75,-20){\large $\tilde{B}$}
\put(-180,55){\large $\Sigma_{xy}$}
\put(-125,-7){ {${0.05}$}}
\put(-93,-7){{${0.2}$}}
\put(-65,-7){{${0.5}$}}
\put(-34,-7){{${5}$}}
\put(-160,110){\large $a)$}
\hspace{1cm}
\includegraphics[width=0.4\textwidth]{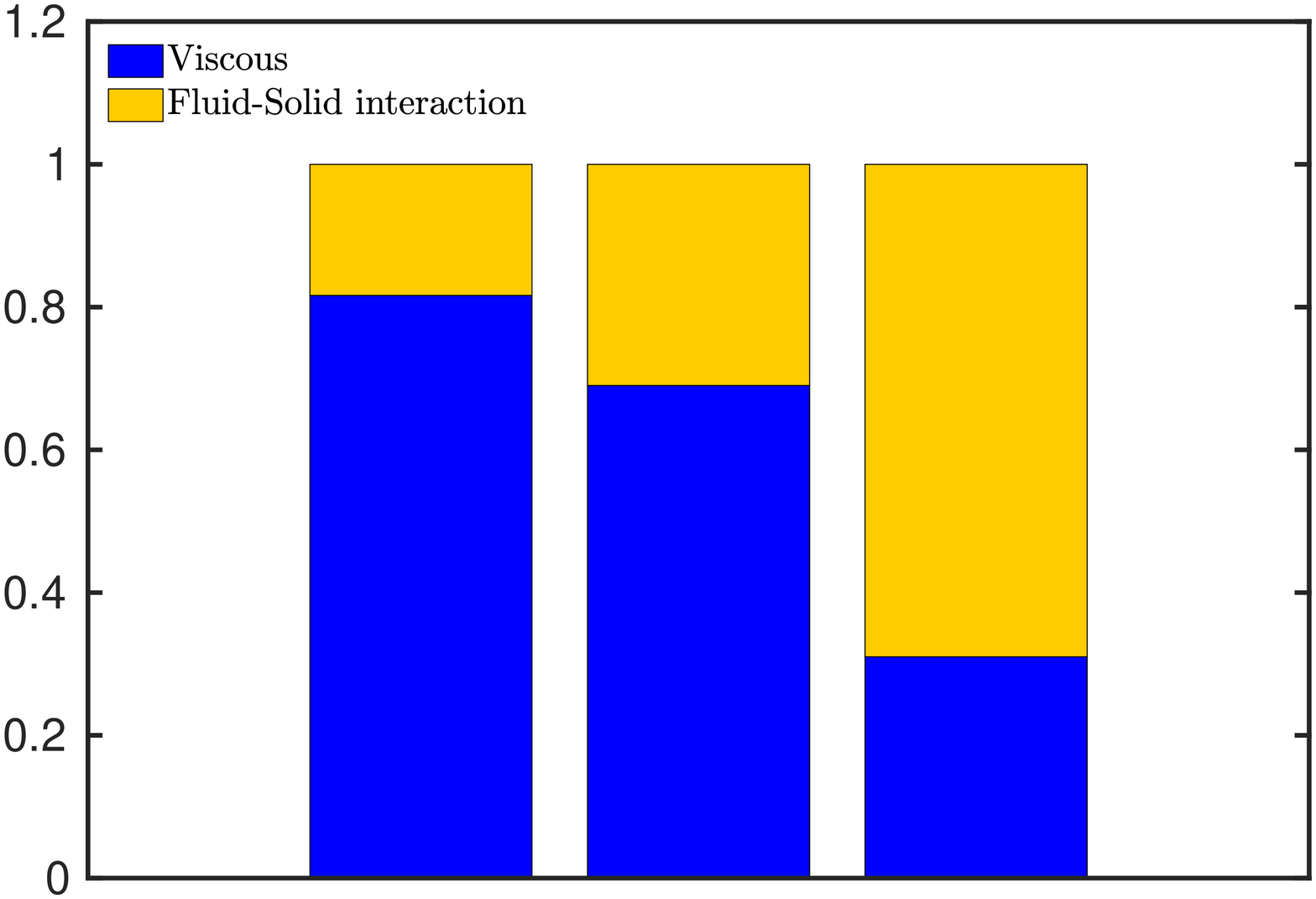}
\put(-75,-20){\large $Re$}
\put(-180,55){\large $\Sigma_{xy}$}
\put(-111,-7){${0.1}$}
\put(-75,-7){${1}$}
\put(-45,-7){${10}$}
\put(-160,110){\large $b)$}
\caption{Stress budget. Relative contributions to the total shear stress from the viscous and fluid-solid interaction stresses versus (a) the filament bending rigidity at a fixed Reynolds number equal to $Re=0.1$ and (b) for different Reynolds numbers and fixed bending rigidity $\tilde{B}=0.5$. The volume fraction is equal to $\phi=0.053$. Note that, the total shear stress is normalised with the total shear stress of each case, cf. figure \ref{fig_visc} for the absolute values.}
\label{fig_budg}
\end{figure}
To better understand the rheological behavior of the suspensions we next examine the different contributions to the total shear stress, as derived in the previous section, see equation (\ref{batchelor1}). Figure \ref{fig_budg}a reports the relative contribution of the viscous and fluid-solid interaction stresses to the total shear stress for the case with low inertia, $Re=0.1$, and for different values of $\tilde{B}$, whereas \ref{fig_budg}b considers the behaviour for different Reynolds numbers at a fixed bending rigidity $\tilde{B}=0.5$. Note that, the Reynolds stresses are negligible also at the highest Reynolds numbers considered and are therefore not displayed in figure \ref{fig_budg}. 

Figure  \ref{fig_budg}a clearly reveals that the contribution to the total stress from the suspended filaments reduces when decreasing the fiber rigidity; on the other hand, the results in figure \ref{fig_budg}b show that changes of the Reynolds number strongly affect the suspension and that the filament contribution increases with inertia and eventually becomes the dominant effect at $Re=10$, where the fluid-solid interaction stress is $66\%$ of the total stress. Thus, we can attribute the large increase of the suspension viscosity observed in figure \ref{fig_visc}b to the fluid-solid interaction forces. Interestingly, the increase of the filament stress with the Reynolds number is more evident from $Re=1$ to $Re=10$ indicating a strong non-linear behaviour, as also seen in the relative viscosity curve. Note that, the same trend is observed also for the other values of $\tilde{B}$ under investigation, with the increase of the fluid-solid interaction term being higher for stiff fibers (data not shown). The increase of the stress component due to the fluid-solid interaction can be related to the increase in the drag force experienced by the filaments at finite inertia, similarly to what observed for cylinders and spheres \citep{fornberg1980numerical}. 

\begin{figure} 
\centering
\includegraphics[width=0.5\textwidth]{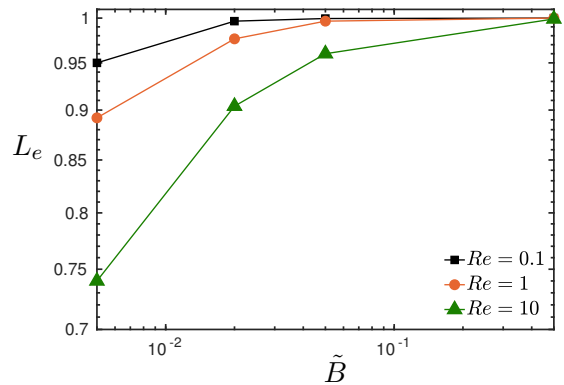}
\put(-90,-10){\large $\tilde{B}$}
\put(-208,76){\large $L_e$}
\caption{Time and ensemble average end-to-end distance $L_e$ of the suspended filaments as a function of their bending rigidity $\tilde{B}$ for the different Reynolds numbers investigated, as indicated in the legend.}
\label{e2e}
\end{figure}
To examine the filament dynamics and their deformation, we now consider the mean distance between the two ends of each filament, where the average is performed over time and the number of filaments. The data pertaining all the different cases under investigation are shown in figure \ref{e2e}. The figure indicates that the end-to-end distance increases as the bending stiffness is increased, i.e.,~the filaments deformationt decreases for larger values of $\tilde{B}$. Moreover, the average end-to-end distance decreases with the Reynolds number, i.e.,~the filaments deform more when increasing the inertial effects. These observations are consistent with the results of the relative viscosity and stress budget discussed above, as we have reported larger filament stresses for the stiff cases and larger suspension viscosity at finite Reynolds number.

\begin{figure} 
\begin{center}
\includegraphics[width=.38\textwidth]{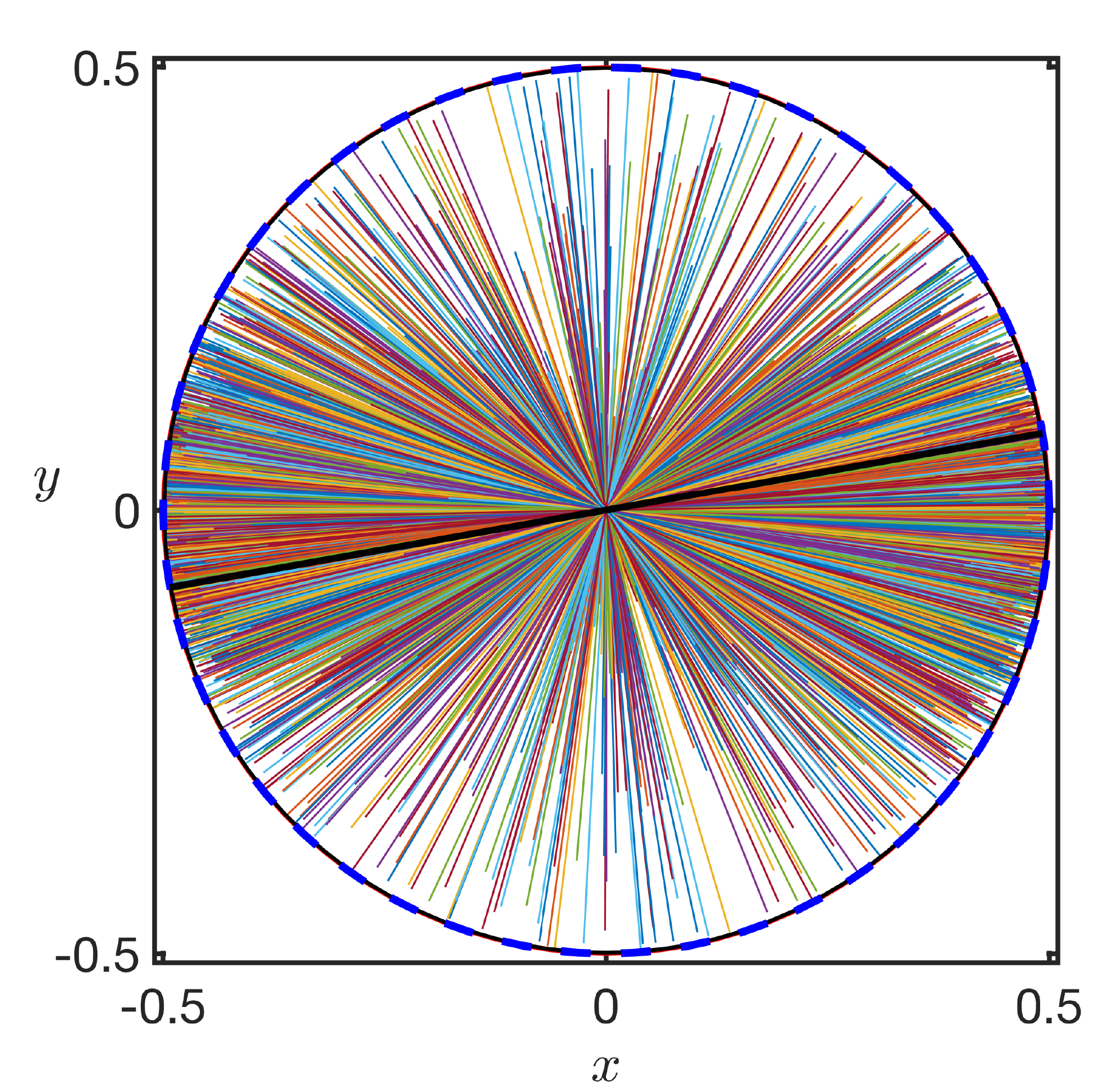} 
\put(-77,-10){\large $(a)$}
\includegraphics[width=.4\textwidth]{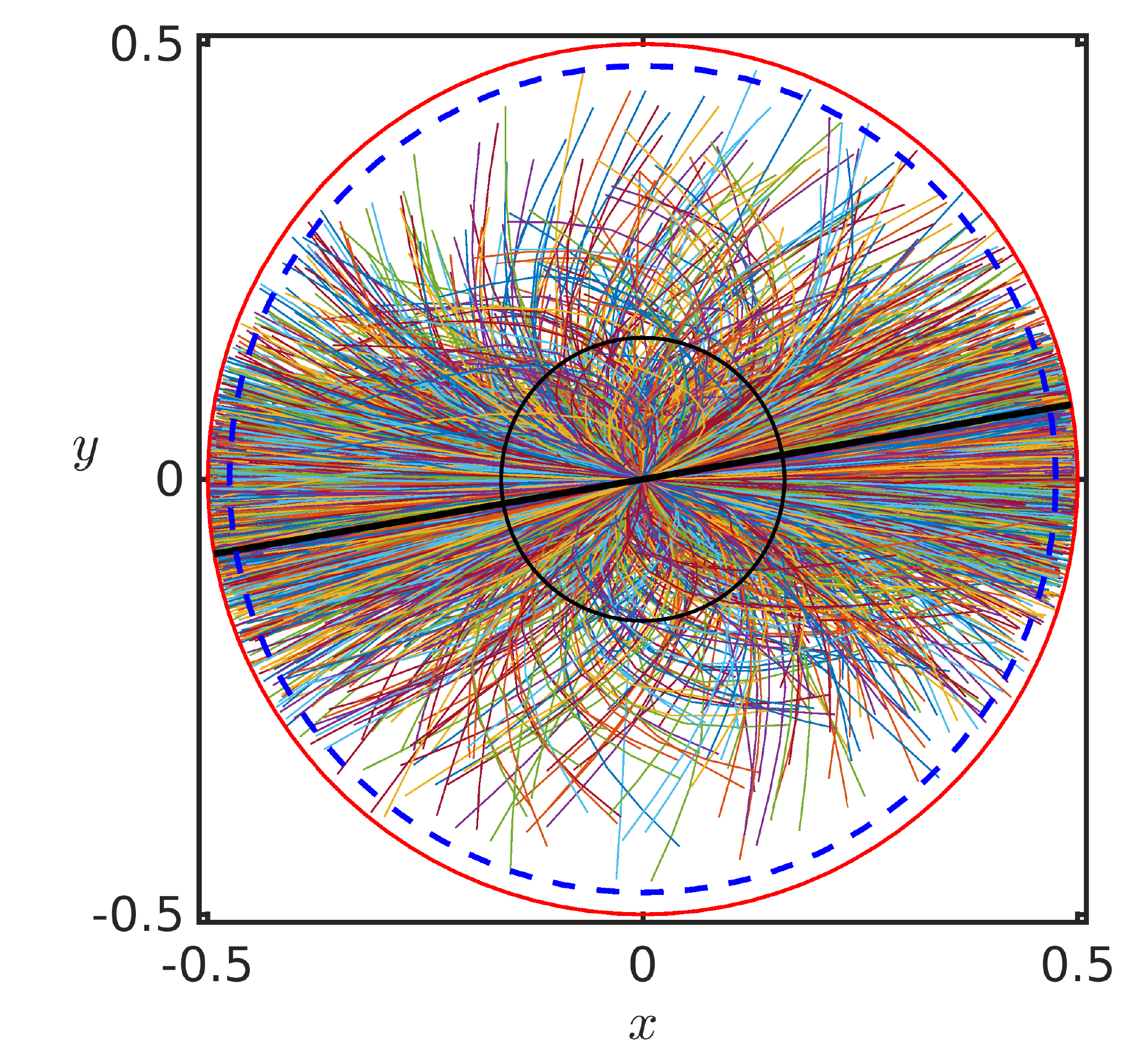}
\put(-75,-10){\large $(b)$}
\newline
\includegraphics[width=.4\textwidth]{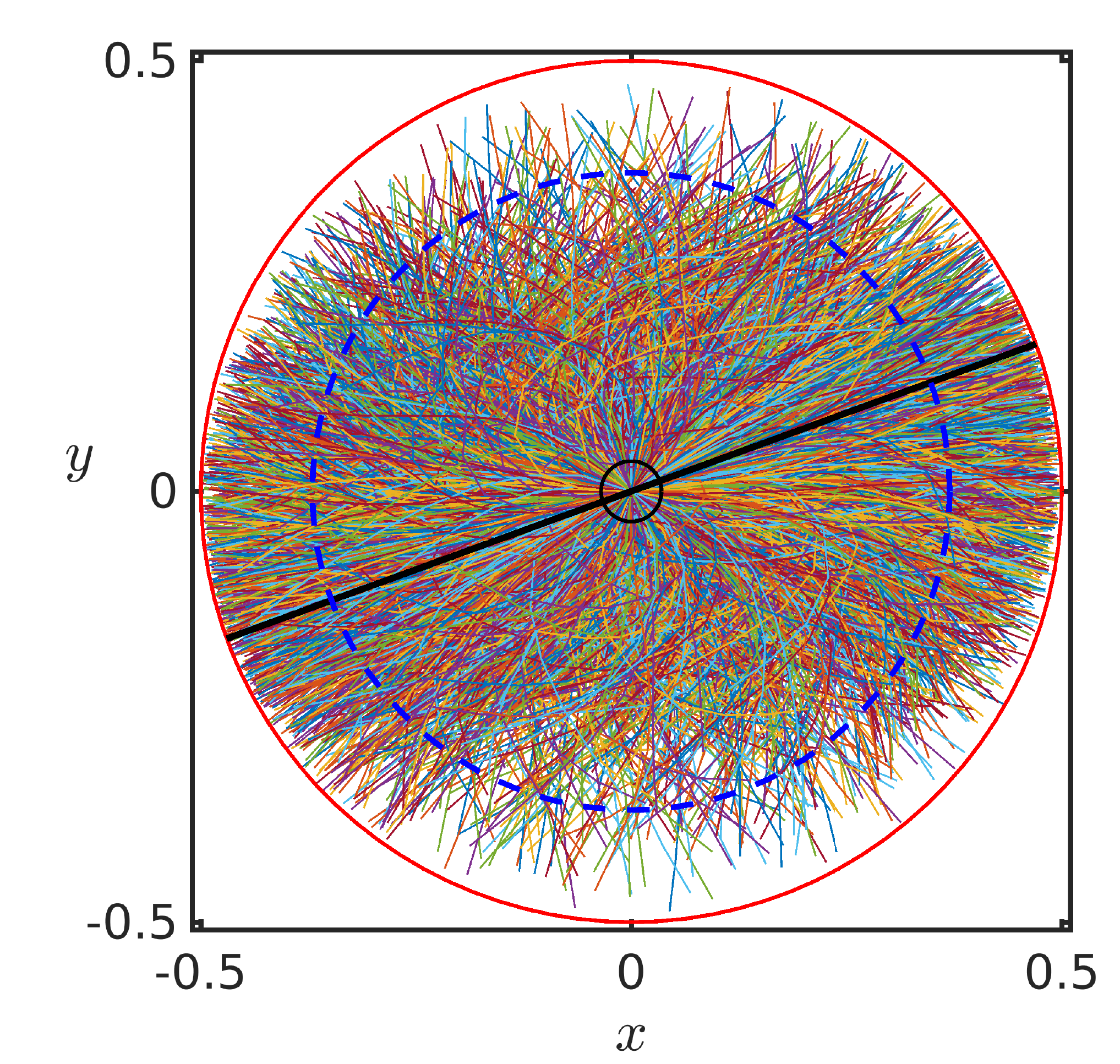}
\put(-75,-10){\large $(c)$}
\end{center}
\caption{Snapshots of co-located filaments projected onto the shear plane in statistically steady state for a) $\tilde{B}=0.5, Re=0.1$ b) $\tilde{B}=0.005, Re=0.1$ and c) $\tilde{B}=0.005, Re=10$. The solid red line is the circle with diameter equal to the filaments length whereas the blue dashed line represents the circle with diameter equal to the mean end-to-end distance reported in figure \ref{e2e}. The black circle represents the minimum end-to-end distance for each case and the black solid lines denote the average orientation with respect to the walls. Note that the three lines coincide in the case of rigid fibers in panel (a).The orientation angles for the cases with flexible fibers are computed on the line connecting the two ends of the filament.}
\label{coloc}
\end{figure}
In order to visualise the filaments deformation in the suspension flow at the statistically steady state, we display them co-located with their center positioned at the origin of the axis, i.e.\ we move their center to (0,0,0) and plot all filaments in the same graph. These are shown in figure \ref{coloc} where we display the projection of the filament configuration in the shear plane for three representative cases: a stiff ($\tilde{B}=0.5$) and a flexible case ($\tilde{B}=0.005$) at negligible inertia (low Reynolds number, $Re=0.1$) and a flexible case at high Reynolds number $\tilde{B}=0.005$ and $Re=10$. Infigure \ref{coloc}, the solid red line is the circle with diameter equal to the filaments length whereas the blue dashed line represents the circle with diameter equal to the mean end-to-end distance reported in figure \ref{e2e}; the black circle represents the minimum end-to-end distance for each case and the solid line the mean orientation of the filaments. For $\tilde{B}=0.5$ and $Re=0.1$ (low inertia and rigid-like filaments) (panel a), the majority of the filaments exhibit negligible bending and indeed the mean end to end distance is very close to the fiber length. Larger deformations are observed for $\tilde{B}=0.005$ and $Re=0.1$ (figure \ref{coloc}b), with a minimum end-to-end distance lower than that in panel a. Finally, for $\tilde{B}=0.005$ and $Re=10$ (figure \ref{coloc}c) the filaments exhibit a substantial bending with a smaller minimum of the instantaneous end-to-end distance. Note also that the filaments deformation is larger for orientation angles around $135^\circ$ and $315^\circ$ with respect to the flow direction i.e. in the compression region of the shear plane, which can be attributed to the maximum acceleration achieved at this inclination and to the buckling of the filaments under compressive forces  \citep{tornberg2004simulating,becker2001instability}. For angles close to $45^\circ$ and $225^\circ$ (the extension region of the shear plane) the acceleration is small as the filaments are aligned to the imposed shear with the hydrodynamic forces working to extend the filaments, which results in the filaments exhibiting small deformations.

\begin{figure} 
\centering
\includegraphics[width=0.5\textwidth]{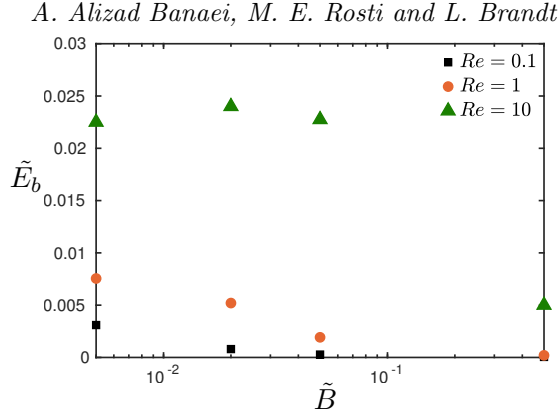}
\put(-90,-10){\large $\tilde{B}$}
\put(-205,74){\large $\tilde{E_b}$}
\caption{Bending energy $\tilde{E_b}$ as a function of the filaments bending stiffness $\tilde{B}$ for the different Reynolds numbers indicated in the legend. }
\label{eng}
\end{figure}
In filament suspensions, there is an exchange between fluid kinetic energy and filament bending energy, where the amount of bending energy directly affects the suspensions bulk elasticity. It is therefore interesting to study the mean filaments bending energy, defined as follows
\begin{equation}\label{bendenergy}
\tilde{E_b}=\frac{1}{n}\tilde{B}\sum\int_0^1 \left |\left |\frac{\partial^2 \bm{X}}{\partial s^2}\right |\right |^2 ds.
\end{equation}
Equation (\ref{bendenergy}) shows that
the energy depends linearly on the bending rigidity $\tilde{B}$ and quadratically on the filament curvature $||\frac{\partial^2 \bm{X}}{\partial s^2}||$; this implies that the bending energy tends to zero for very soft ($\tilde{B}\to 0$) and very stiff ($||\frac{\partial^2 \bm{X}}{\partial s^2}|| \to 0$) cases and a maximum should exist for intermediate values of $\tilde{B}$. The bending energy of all the different suspensions under consideration is displayed in figure \ref{eng}: the peak corresponding to the maximum elastic energy of the suspension shifts to low $\tilde{B}$ when the Reynolds number decreases and, as expected, the bending energy approaches zero for the most stiff cases. It can also be inferred from the figure that for a fixed rigidity, the bending energy increases with the Reynolds number, especially from $Re=1$ to $Re=10$ due to the larger filament deformation.

\begin{figure} 
\centering
\includegraphics[width=0.5\textwidth]{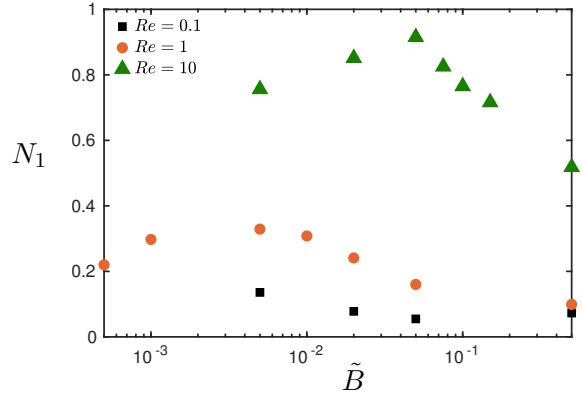}
\put(-90,-10){\large $\tilde{B}$}
\put(-215,76){\large $N_1$}
\caption{First normal stress difference $N_1$ as a function of the bending rigidity $\tilde{B}$ for the different Reynolds numbers under investigation. The volume fraction is fixed to $\phi=0.053$.}
\label{normal}
\end{figure}
As mentioned before, the bending energy can be directly related to the visco-elastic behavior of the filament suspension. In order to quantify this effect, we examine the first normal stress difference for the different Reynolds numbers and bending rigidities under investigation, see figure \ref{normal}. As expected for a visco-elastic suspension, the first normal stress difference is positive, similarly to what found in the case of polymers, deformable particles and capsules \citep{mewis_wagner_2012a,matsunaga2016rheology,rosti2018suspensions,shahmardi_zade_ardekani_poole_lundell_rosti_brandt_2019a}. Interestingly, the trend in figure \ref{normal} is similar to that of the bending energy, with  the first normal stress difference increasing with the flow inertia; furthermore, we note the presence of a maximum of $N_1$, whose location is shifting to larger values of $\tilde{B}$ when the Reynolds number is increased. In order to capture accurately the location of the maximum, we performed additional simulations for $Re=1$ and $10$.. The filament suspension exhibits the highest first normal stress difference at around $\tilde{B}=0,05$ for $Re=10$ and around $\tilde{B}=0,005$ for $Re=1$, thus suggesting that the value of $\tilde{B}$ for which the maximum first normal stress difference is achieved scales approximately with the Reynolds number. Note that, for $Re=0.1$ the peak of the first normal stress difference is expected for values of $\tilde{B}$ below those considered here.
 
\begin{figure} 
\centering
\includegraphics[width=0.45\textwidth]{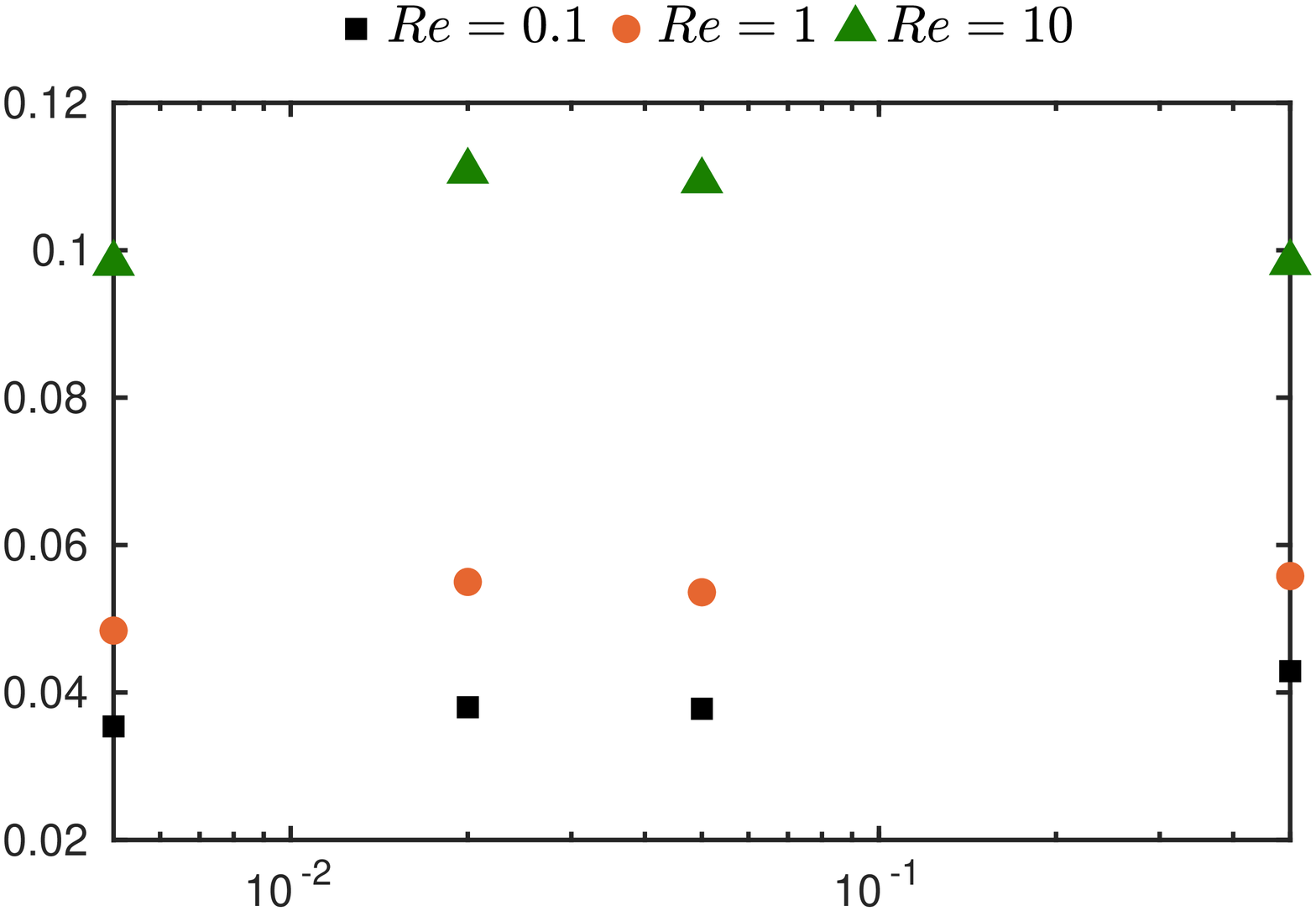} 
\put(-83,-7){\large $\tilde{B}$}
\put(-189,60){\large $u'$}
\put(-189,115){\large $a)$}
\hspace{1cm}
\includegraphics[width=0.45\textwidth]{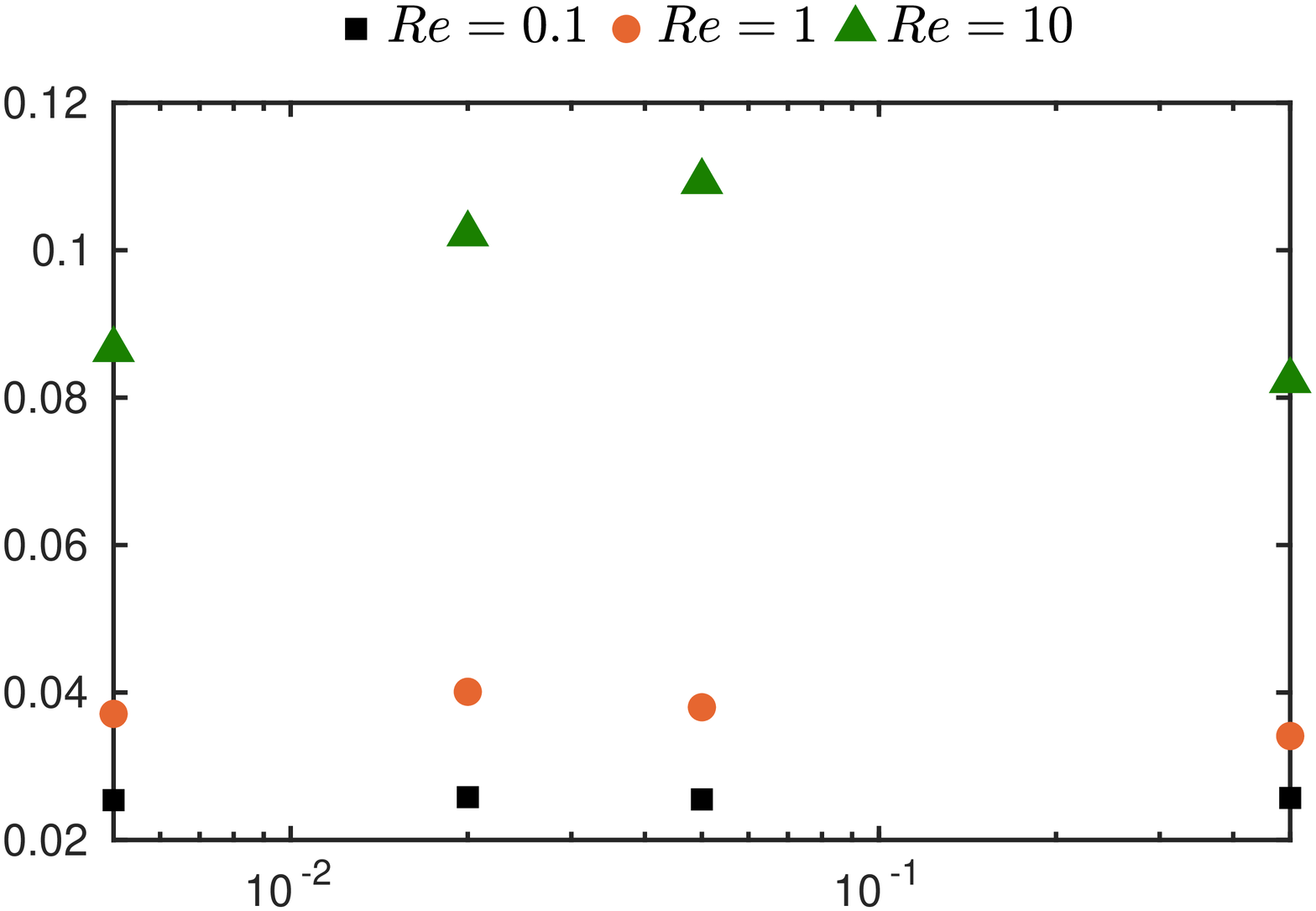}
\put(-83,-7){\large $\tilde{B}$}
\put(-185,115){\large $b)$}
\put(-189,60){\large $v'$}
\caption{Root mean square of the streamwise $u'$ and wall normal $v'$ velocity fluctuations integrated across the channel width as a function of the filament bending stiffness $\tilde{B}$  for different values of the Reynolds number $Re$, as indicated in the figure.}
\label{vi}
\end{figure}
Before concluding this section, we consider the root mean square of the fluid velocity fluctuations induced by the presence of the filament. The integral across the channel of the streamwise and wall-normal velocity fluctuations are displayed in figure \ref{vi}. As expected, we note that the velocity fluctuations increase with the Reynolds number, due to larger fluid-solid interaction forces as $Re$ increases. The magnitude of these fluctuations is approximately independent of the bending rigidity $\tilde{B}$ for the different Reynolds numbers examined, except for a small peak in correspondence to the maximum of the first normal stress difference, which is more evident at finite inertia. This weak dependency on the rigidity suggests that, for the parameters considered in this study, fluid velocity fluctuations are mainly induced by the filament rotation.

\subsection{Rheology varying the filament volume fraction}
In this section we investigate the effect of the filament volume fraction at $Re=0.1$ and $Re=10$ for two different values of rigidity, $\tilde{B}=0.5$ and $\tilde{B}=0.02$. The former case is chosen as representative of the behavior of rigid fibers as the deformation is negligible also at the highest Reynolds number  considered.

\begin{figure} 
\centering
\includegraphics[width=0.45\textwidth]{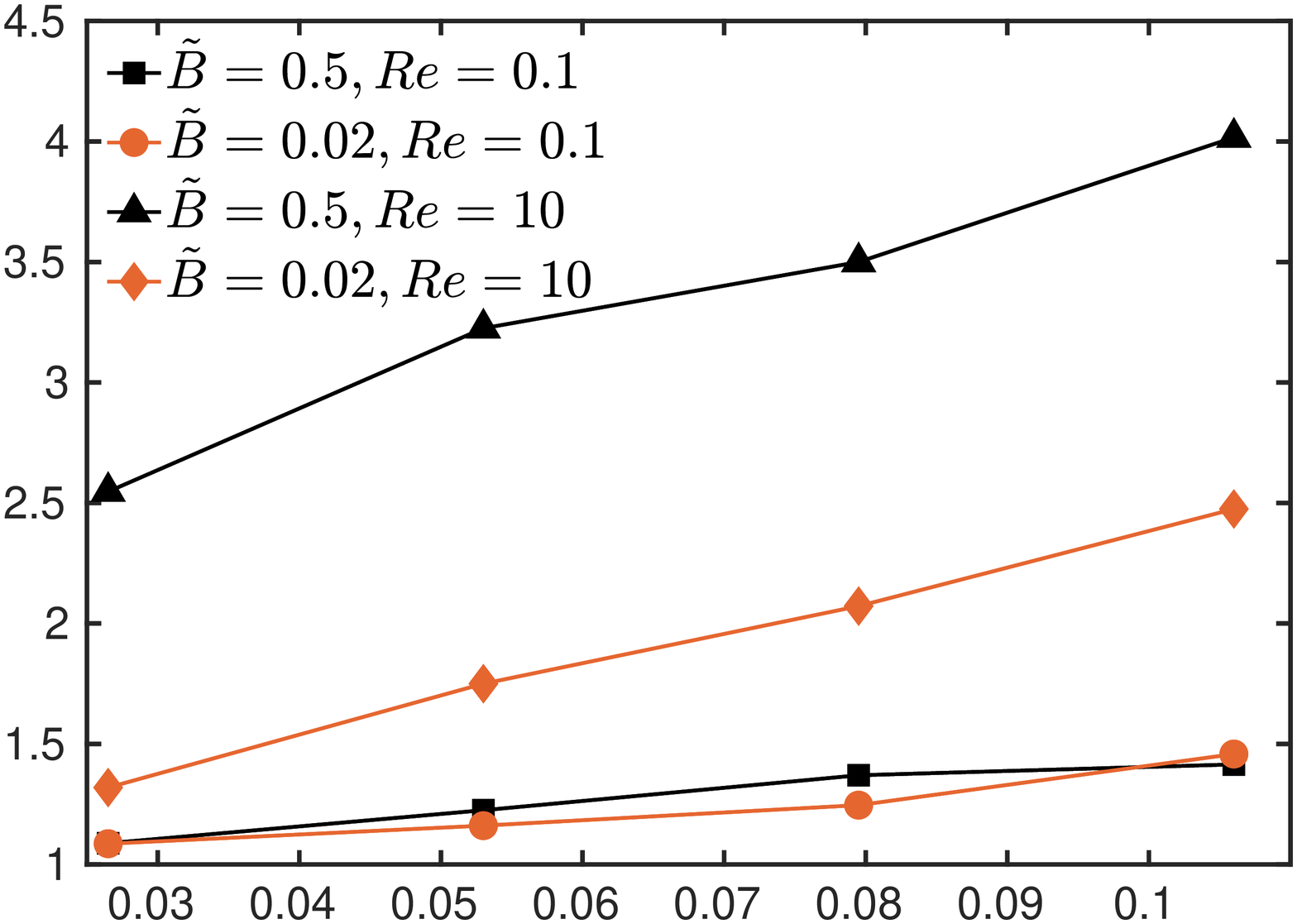} 
\put(-80,-10){\large $\phi$}
\put(-185,60){\large $\eta$}
\put(-185,115){\large $a)$}
\hspace{1cm}
\includegraphics[width=0.45\textwidth]{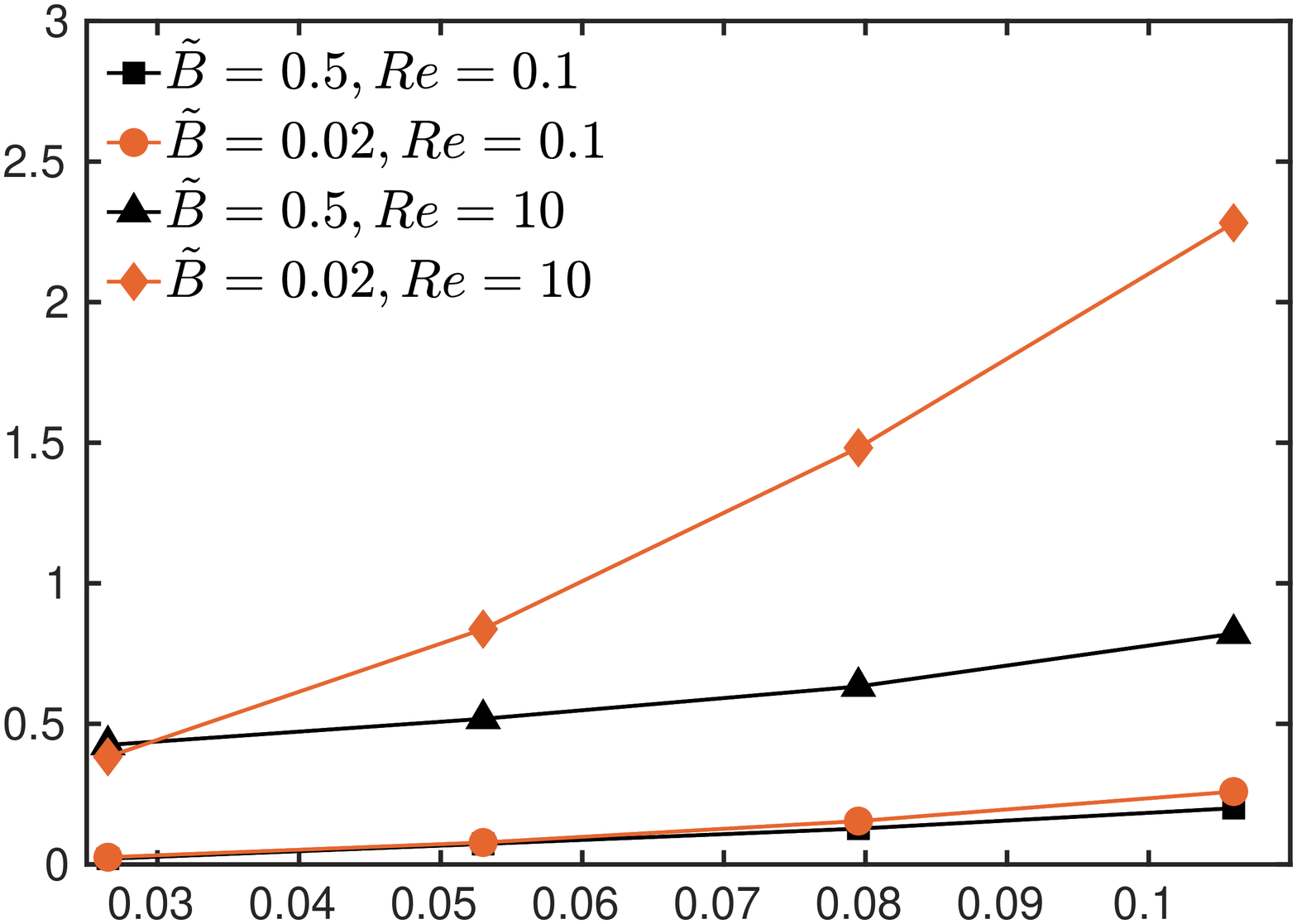}
\put(-185,115){\large $b)$}
\put(-80,-10){\large $\phi$}
\put(-189,60){\large $N_1$}
\caption{Relative viscosity and first normal stress difference as a function of the filament volume fraction for different Reynolds numbers and flexibilities as indicated in the legend. }
\label{viscvf}
\end{figure}
\begin{figure} 
\centering
\includegraphics[width=.6\textwidth]{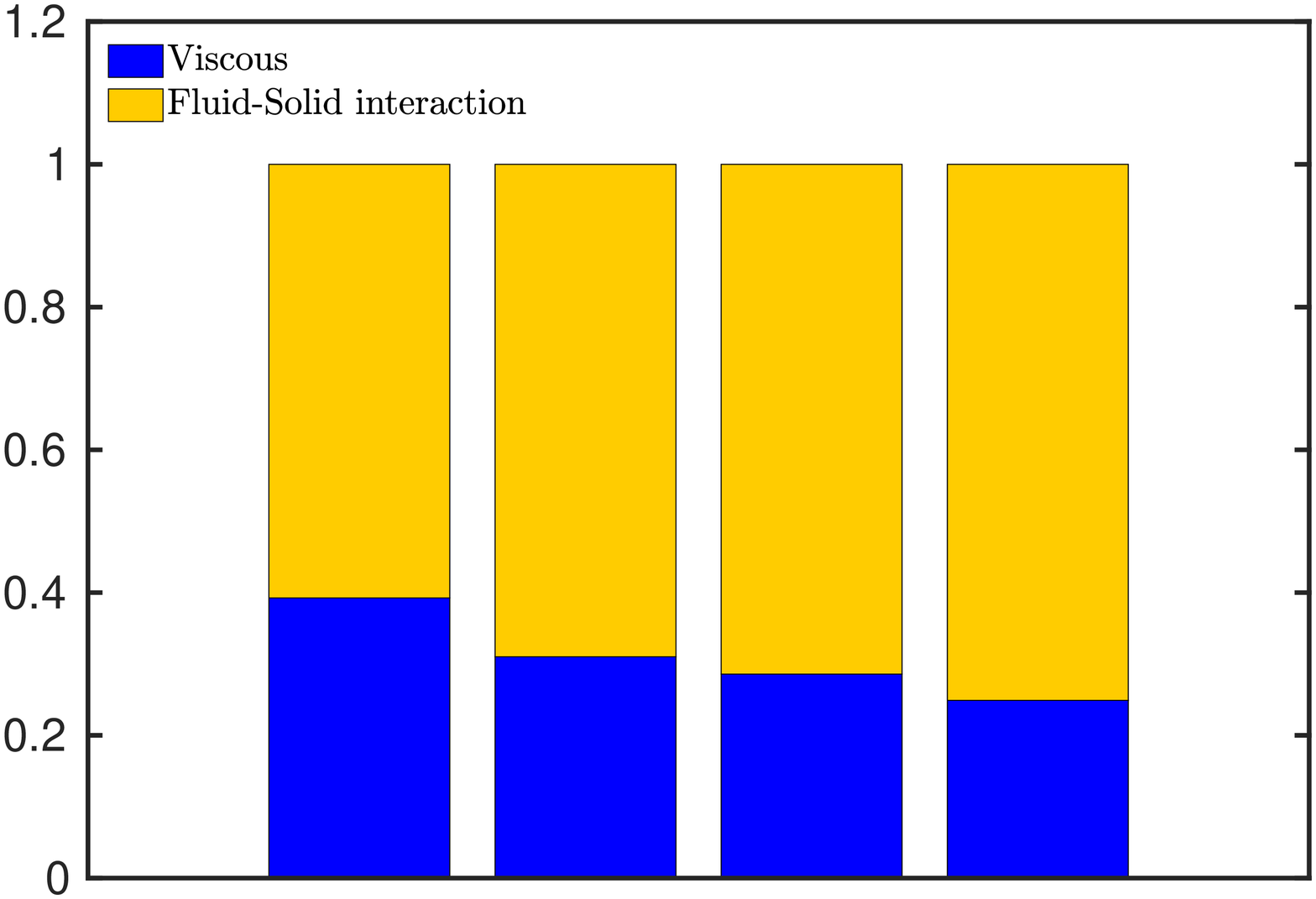}
\put(-115,-20){\large $\phi$}
\put(-255,75){\large $\Sigma_{xy}$}
\put(-115,-35){\large $(a)$}
\put(-183,-7){${0.0265}$}
\put(-140,-7){${0.053}$}
\put(-101,-7){${0.0795}$}
\put(-60,-7){${0.106}$} \\
\includegraphics[width=.6\textwidth]{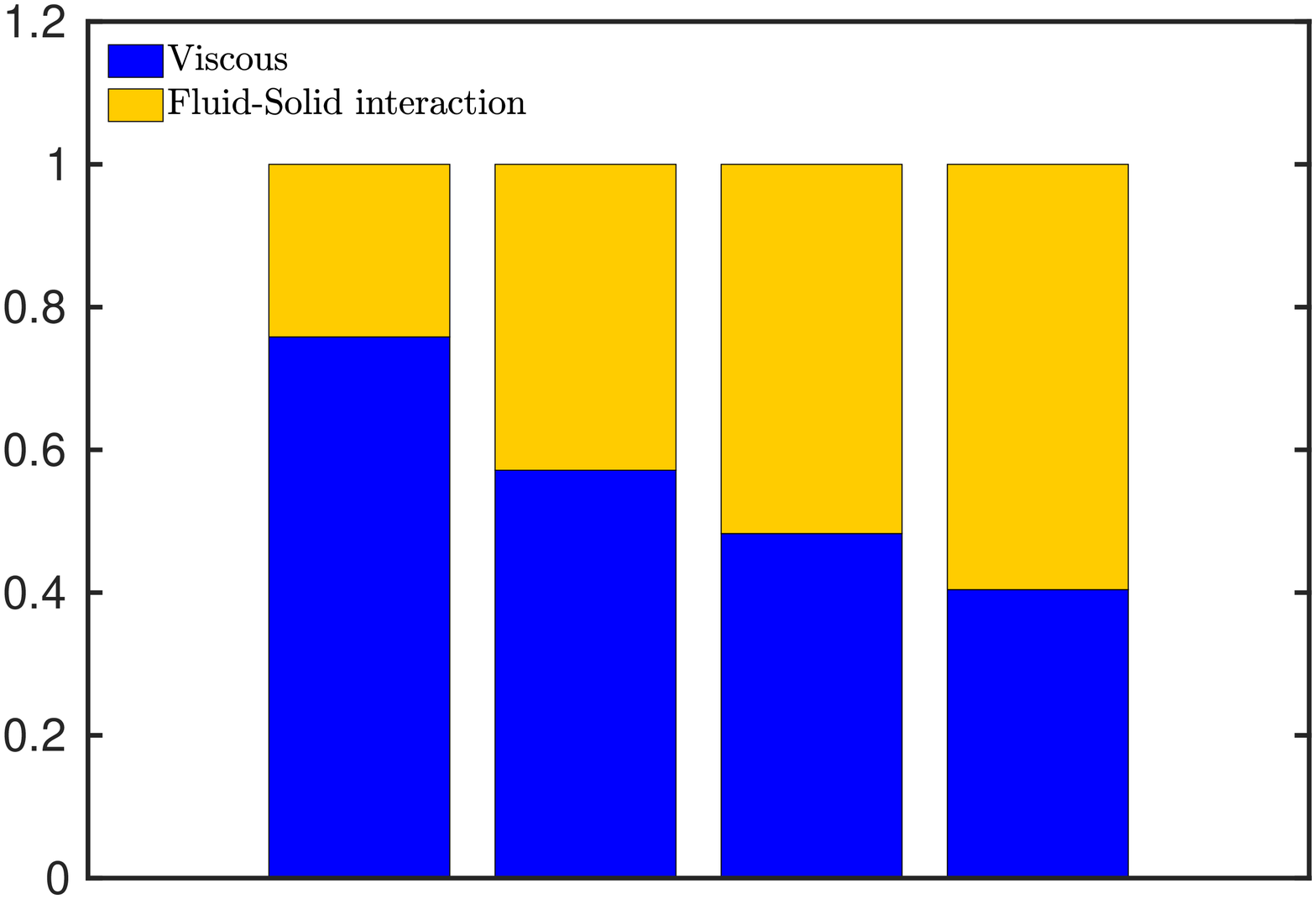}
\put(-115,-20){\large $\phi$}
\put(-255,75){\large $\Sigma_{xy}$}
\put(-115,-35){\large $(b)$}
\put(-183,-7){${0.0265}$}
\put(-140,-7){${0.053}$}
\put(-101,-7){${0.0795}$}
\put(-60,-7){${0.106}$}
\caption{Stress budget. Relative contributions to the total shear stress from the viscous and fluid-solid interaction stresses versus the filament volume fraction at $Re=10$, for a) $\tilde{B}=0.5$ and b) $\tilde{B}=0.02$. Note that, the total shear stress is normalised with the total shear stress of each case, cf. figure \ref{viscvf} for the absolute values.}
\label{budvf}
\end{figure}
In figure \ref{viscvf} we present the suspension rheological behavior in terms of relative viscosity and first normal stress difference. The data infigure \ref{viscvf} clearly show that the relative viscosity increases with the volume fraction except for the cases at the two highest $\phi$ for negligible inertia and rigid filaments, $Re=0.1$ and $\tilde{B}=0.5$. In this case, we report  a slight decrease of the effective viscosity which we explain by the orientation angle of the filaments. Indeed, by increasing the volume fraction from $7.9\%$ to $10.6\%$, filaments become more aligned with the mean flow which results in lower viscosity and also reduced normal stress difference. A similar reduction has also been observed by \cite{lindstrom2008simulation}. Furthermore, we note only a small difference in the relative viscosity for the two cases with different bending rigidities at negligible inertia, $Re=0.1$, which can be attributed to the small difference in the mean filament deformation, as will be discussed later. At finite inertia, $Re=10$, the suspensions of more flexible filaments display lower viscosity, confirming the decrease with the flexibility discussed above.

The first normal stress difference, see panel b of figure \ref{viscvf}, also increases with the volume fraction, more visibly at the highest Reynolds number considered, $Re=10$. The first normal stress difference is larger for the suspension of most flexible filaments at higher Reynolds numbers, while it increases slowly at larger $\phi$ and $\tilde{B}=0.5$. At low Reynolds numbers, the values of $N_1$ are similar for the suspensions of rigid and deformable filament, and a difference is only visible at the highest $\phi$ considered, which can be related to the formation of a micro-structure in the suspensionsso that the filaments have lower mobility as further discussed below.

The stress budget pertaining the cases at different volume fraction $\phi$ is displayed in figure \ref{budvf} where we report the relative contribution from viscous and elastic stresses (the absolute values of the shear viscosity being depicted in figure \ref{viscvf}). As expected from the results above, the contribution from the fluid-solid interaction force increases with the filament rigidity and volume fraction.

\begin{figure} 
\centering
\includegraphics[width=0.45\textwidth]{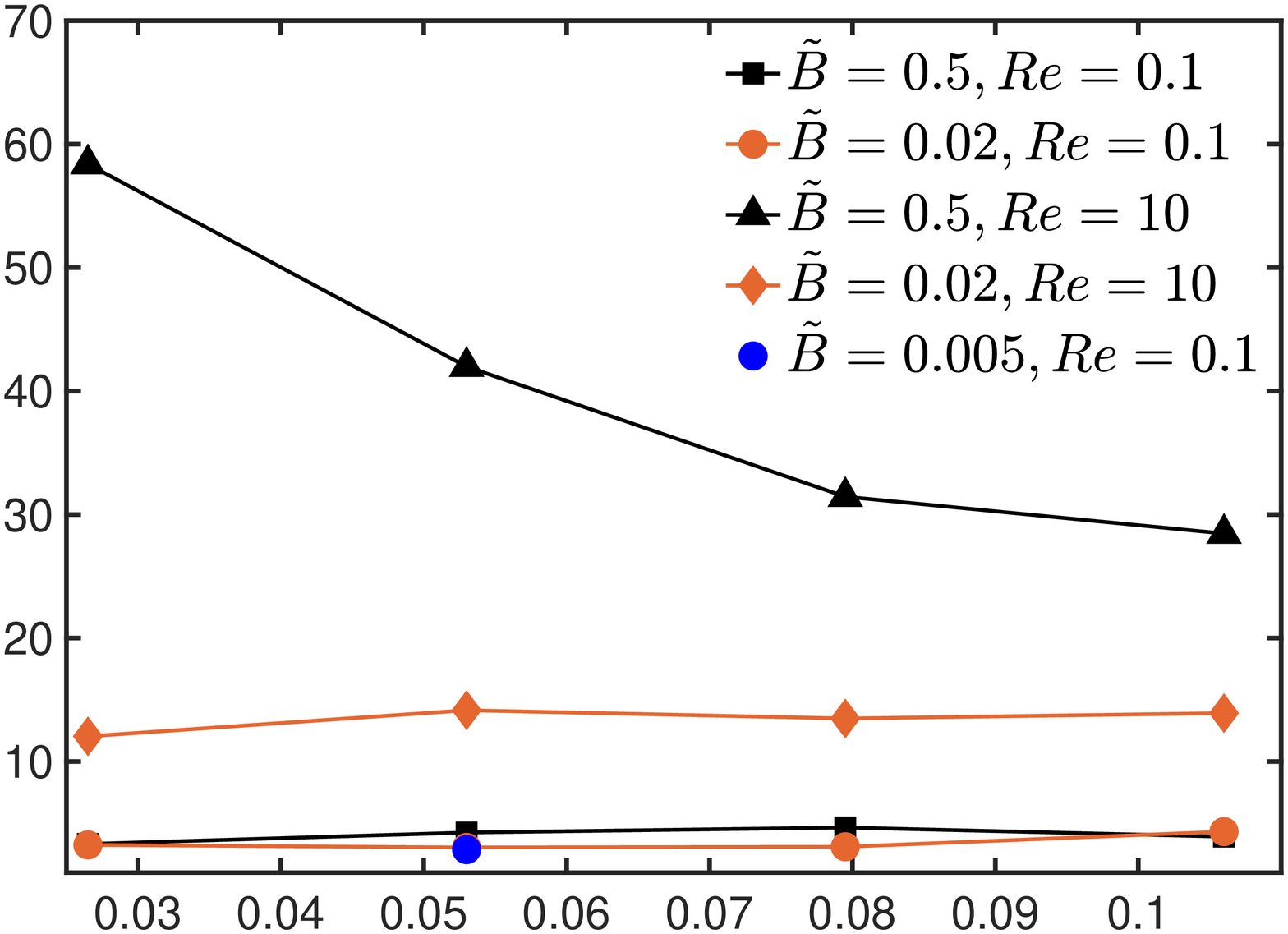} 
\put(-85,-12){\large $\phi$}
\put(-195,65){\large $\frac{\bar{\Sigma}^f_{xy}}{\phi}$}
\put(-185,115){\large $a)$}
\hspace{1cm}
\includegraphics[width=0.45\textwidth]{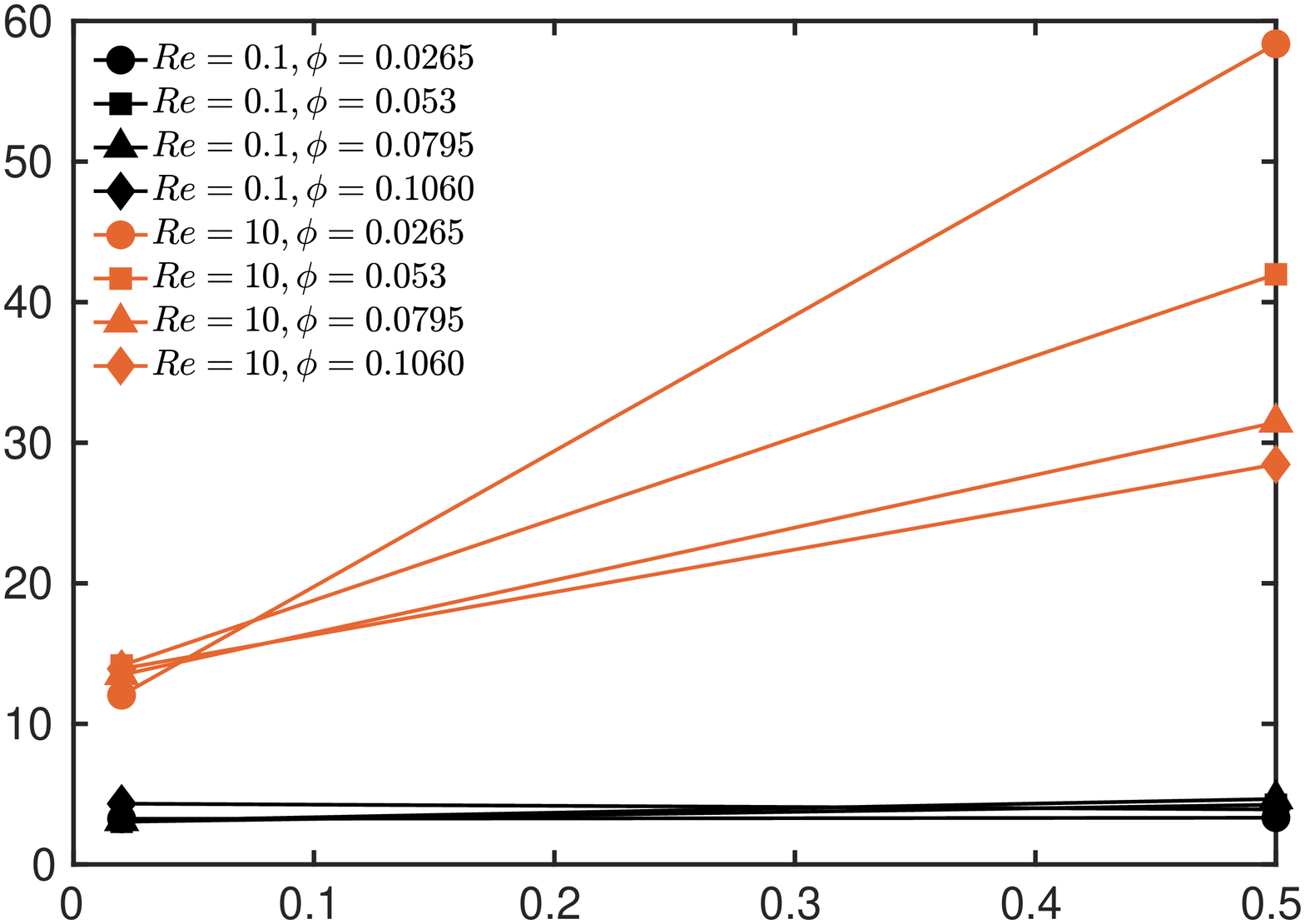}
\put(-85,-12){\large $\tilde{B}$}
\put(-185,115){\large $b)$}
\put(-195,65){\large $\frac{\bar{\Sigma}^f_{xy}}{\phi}$}
\caption{Filament contribution to the total shear stress scaled with the volume fraction as a function of (a) volume fraction and (b) bending stiffness}
\label{vf_sc}
\end{figure}

\begin{figure} 
\centering
\includegraphics[width=0.45\textwidth]{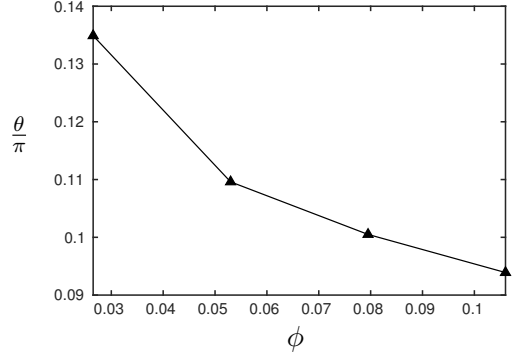} 
\put(-85,-12){\large $\phi$}
\put(-190,65){\large $\frac{\theta}{\pi}$}
\caption{Average orientation angle with the wall for the most stiff filaments in suspensions with $Re=10$ and $\tilde{B}=0.5$.}
\label{orientation}
\end{figure}
Figure \ref{vf_sc} represents the variation of the filament contribution to the average total shear stress scaled by the volume fraction. Note that this contribution includes also lubrication and collision forces. It can be observed in figure \ref{vf_sc}(a) that at low Reynolds number, the contribution is proportional to the volume fraction both for rigid and flexible filaments, meaning that filament-filament interactions are negligible and that deformation is weak to give a visible effect. The blue dot  in figure \ref{vf_sc}(a), pertaining the suspensions with most flexible filaments at $\phi=5.3\%$ (see previous section), also suggests that the filament stress decreases for more deformable objects, although this effect is still small for the cases considered here at negligible inertia. When inertia is important, $Re=10$, the filament stress contribution increases, as shown by the global shear viscosity in figure \ref{viscvf}(a), while the linear dependence on the volume fraction is only observed for the most deformable filaments. On the contrary, the contribution decreases with the volume fraction for the rigid filaments; this is due to the fact that rigid filaments tend to align in the shear direction (see figure \ref{orientation}), which reduces the importance of the short-range interactions. In this case, the filaments move almost as an aggregate, with decreased relative motion. The same data are depicted in figure \ref{vf_sc}(b), as a function of the bending stiffness. At $Re=0.1$, the filament stress is almost constant when changing the bending stiffness whereas at $Re=10$ we clearly observe a decrease of the viscosity with the deformability, which significantly increases with the volume fraction of the filaments;  this is attributed to the combination of decreased deformation as observed for other deformable objects, and to the formation of a more ordered microstructure.

\begin{figure} 
\centering
\includegraphics[width=0.5\textwidth]{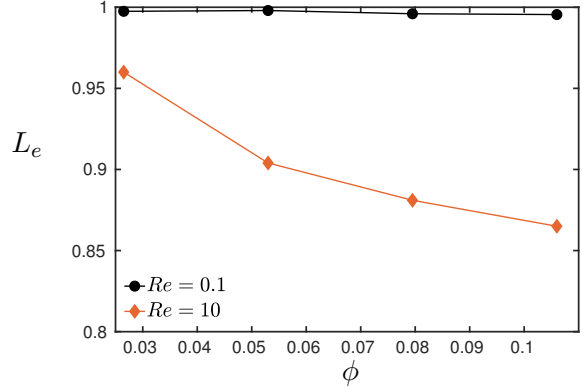}
\put(-90,-10){\large $\phi$}
\put(-215,76){\large $L_e$}
\caption{Average end-to-end distance $L_e$ of the suspended filaments as a function of filament volume fraction, $\tilde{B}=0.02$}
\label{e2evf}
\end{figure}
To quantify the increase of the filament deformation with the Reynolds number and volume fraction, we display  in figure \ref{e2evf} the mean end-to-end distance for the cases with flexible filaments, $\tilde{B}=0.02$. The larger deformations observed at higher volume fraction and Reynolds number can be attributed to increased filament-filament interactions, as opposed to the case of rigid filament just discussed where the interactions are reduced by the formation of an ordered arrangement. Note, again, that for the cases with $\tilde{B}=0.5$ the end-to-end distance is very close to $1$ (filament initial length) for all cases considered and the filaments behave like rigid rods.

\begin{figure} 
\centering
\includegraphics[width=0.45\textwidth]{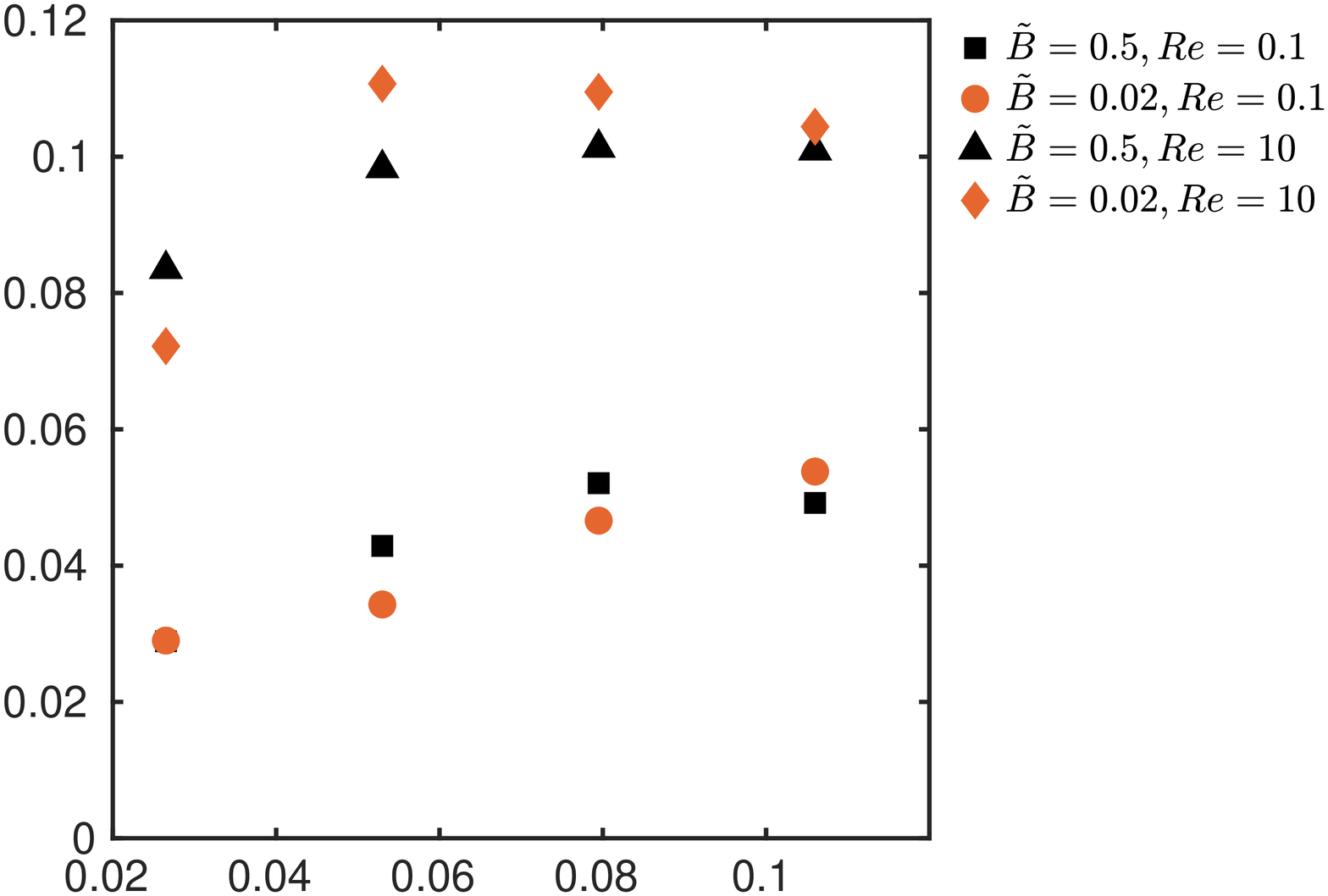} 
\put(-110,-12){\large $\phi$}
\put(-185,65){\large $u'$}
\put(-185,115){\large $a)$}
\hspace{1cm}
\includegraphics[width=0.45\textwidth]{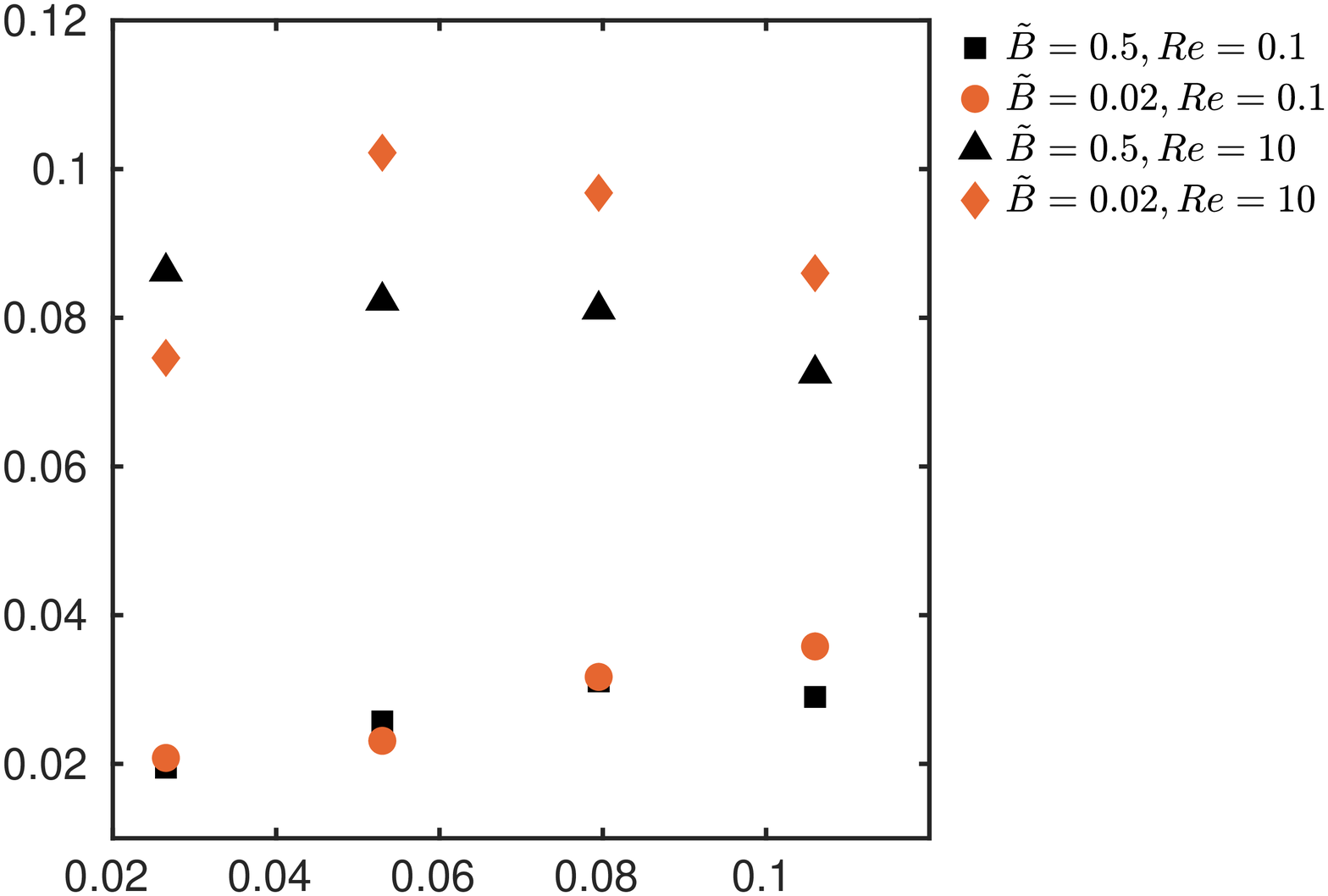}
\put(-110,-12){\large $\phi$}
\put(-185,115){\large $b)$}
\put(-185,65){\large $v'$}
\caption{Root mean square of the streamwise $u'$ and wall normal $v'$ velocity fluctuations integrated across the channel width as a function of the filament volume fraction. The symbol $\circ$ denote $\tilde{B}=0.02, Re=0.1$, $\square$ $\tilde{B}=0.5, Re=0.1$, $\diamond$ $\tilde{B}=0.02, Re=10$ and $\triangle$ $\tilde{B}=0.5, Re=10$.}
\label{viv}
\end{figure}
Finally, the R.M.S. of the streamwise and wall normal velocity fluctuations are depicted in figure \ref{viv} as a function of filament volume fraction for the different cases under investigation. At negligible inertia, the data show that the velocity fluctuations increase with the volume fraction, which is due to the increase of the filament-filament interactions, in agreement with the observations just made about the behavior of the filament stress. At finite inertia, on the other hand, the level of fluctuations first increases and reaches a maximum once the filaments have less freedom to move in the fluid: in the case of rigid filaments, this is due to the formation of an ordered structure, whereas in the case of flexible filaments this can be related to an increased deformation and a reduced effect on the fluid. 

\section{Conclusions}
We have reported the results of numerical simulations of semi-dilute and concentrated filament suspensions for different bending stiffness and Reynolds number. The filaments are modelled as  one-dimensional inextensible slender bodies with fixed aspect ratio ${1}/{16}$ obeying the Euler-Bernoulli beam equation, which enables us to accurately capture the local deformation and curvature of the suspended filaments. The immersed boundary method is used to couple the fluid and solid motion. The code has been validated for three different cases: single filament oscillation due to gravity, single rigid filament rotation in shear flow and filament oscillations in an oscillatory flow, as well as against numerical and experimental results pertaining suspensions of rigid fibers. We therefore move on to study the effect of bending rigidity, Reynolds number and volume fraction on suspension rheology, and analyze the results in terms of stress budget, filament deformation and orientation.

First, at fixed volume fraction, we observe that the relative viscosity of filament suspensions decreases with flexibility, as observed in previous studies and for other flexible objects, e.g.\ capsules, red blood cells and deformable particles, with this reduction with flexibility enhanced at finite inertia. The relative viscosity grows when increasing the Reynolds number due to the larger contribution of the fluid-solid interaction stress to the total stress. The first normal stress difference is positive as in polymeric fluids, and increases with the Reynolds number. Noteworthy, it has a peak for a certain value of the filament bending stiffness, which varies with the Reynolds number, moving towards more rigid suspensions when increasing inertia. This may be related to a resonance condition between flow and filament time scales, as suggested by the behavior of the bending energy \cite[see also][]{rosti2018flexible}. The average end-to-end distance decreases by increasing Reynolds number and decreasing bending rigidity showing that the filaments exhibit larger deformation at higher Reynolds numbers and lower bending rigidities. 

When increasing the filament volume fraction, we observe that the viscosity increases, except for stiff filaments at negligible inertia where we have a clear saturation. The reduction of the effective viscosity with the flexibility mentioned above is more clear at high Reynolds, when filament deform more. On the other hand, it is also stronger at lower volume fraction and decreases as $\phi$ increases. This is due to the formation of a more ordered structure in the flow, where the filaments tend to be more aligned and move as an aggregate, which reduces the filament-filament interactions. Interestingly, the fluid fluctuations first display a maximum at intermediate volume fractions and decrease at the highest considered here, which we explain as a combination of two effects. In the case of rigid filaments, this is due to the formation of an ordered structure at high $\phi$, whereas in the case of flexible filaments this is attributed to an increased deformation which implies a reduced effect on the fluid. It is also interesting to note that although the Reynolds stresses are negligible, the rheological behavior of the suspension is clearly modified at finite inertia by an alteration of the micro-structure.

The present study introduces an approach to investigate filament suspensions in a number of configurations. As an example, our numerical method is also capable to study suspensions with a distribution in filament lengths as often found in experimental configurations. Also, a positive first normal stress difference as in polymeric fluids suggests the idea to study the behavior of finite-size flexible fibers in turbulent flows. These results also show the importance of the short-range interactions among filaments if one wishes to study suspensions at higher volume fractions. In this case, a more accurate modelling of friction and contact forces becomes fundamental to properly capture the global system behavior. In this framework, multiscale approaches might be a viable approach.

\section*{Acknowledgements}
This work was supported by the European Research Council grant no. ERC-2013- CoG-616186, TRITOS. The authors acknowledge computer time provided by SNIC (Swedish National Infrastructure for Computing) and the support from the COST Action MP1305: Flowing matter.

\bibliographystyle{jfm}

\end{document}